\documentclass[toc]{PoS}

\title{Hard-scattering and Jets from RHIC to LHC: a critical review.}

\ShortTitle{Hard-scattering and Jets: a critical review.}

\author{\speaker{Michael J. Tannenbaum}%
\thanks{Supported by the U.S. Department of Energy, Contract No. DE-AC02-98CH1-886.}\\
        Physics Dept., 510c, Brookhaven National Laboratory, Upton, NY 11973-5000, USA\\
        E-mail: \email{mjt@bnl.gov}}

\abstract{Jets in hadron collisions are very complicated with a long learning curve replete with errors. In relativistic heavy ion (RHI) collisions, it is  likely that jets will be much more complicated with an even longer and more difficult learning curve. Hard scattering is more easily observed via single particle and few particle correlation measurements. The main advantage of jets is higher rate at large $p_T$, plus the possibility of detailed studies of soft fragmentation if the soft fragments can be separated from the background. A critical review of the possibility of using jets as a probe of hard-scattering in RHI collisions is presented along with other probes and measurements which the author considers much more likely to reveal the interesting physics in Pb+Pb collisions at the LHC. Finally, a list of unanswered questions raised by results at RHIC is presented.}

\FullConference{High-pT physics at LHC--- \\
		 March 23-27 2007\\
		 University of Jyv\"askyl\"a, Finland}

\def\lsim{\raise0.3ex\hbox{$<$\kern-0.75em\raise-1.1ex\hbox{$\sim$}}}
\def\gsim{\raise0.3ex\hbox{$>$\kern-0.75em\raise-1.1ex\hbox{$\sim$}}}

\def\mean#1{\left<#1\right>}
\def\Journal#1#2#3#4{ {\it{#1}} {\bf #2}, #3 (#4)}
\def\IJMPA{{Int. J. Mod. Phys.}~{\rm A}}
\def\IJMPE{{Int. J. Mod. Phys.}~{\rm E}}

\def\JP{{J. Phys.}}
\def\JPG{{J. Phys.}~{\rm G}}
\def\JPCS{{J. Phys: Conf. Series\ }}

\def\NPA{{Nucl. Phys.}~{\rm A}}
\def\NPB{{Nucl. Phys.}~{\rm B}}
\def\PLB{{Phys. Lett.}~{\rm B}}

\def\PRL{Phys. Rev. Lett.\ }
\def\PRD{{Phys. Rev.}~{\rm D}}
\def\PRC{{Phys. Rev.}~{\rm C}}
\def\PPNP{{Prog. Part. Nucl. Phys.}}
\def\ZPC{{Z. Phys.}~{\rm C}}

\begin{document}

\section{Soft and Hard Scattering in p-p collisions}
\begin{figure}[!thb]
\begin{center}
\begin{tabular}{cc}
\includegraphics[width=0.50\linewidth]{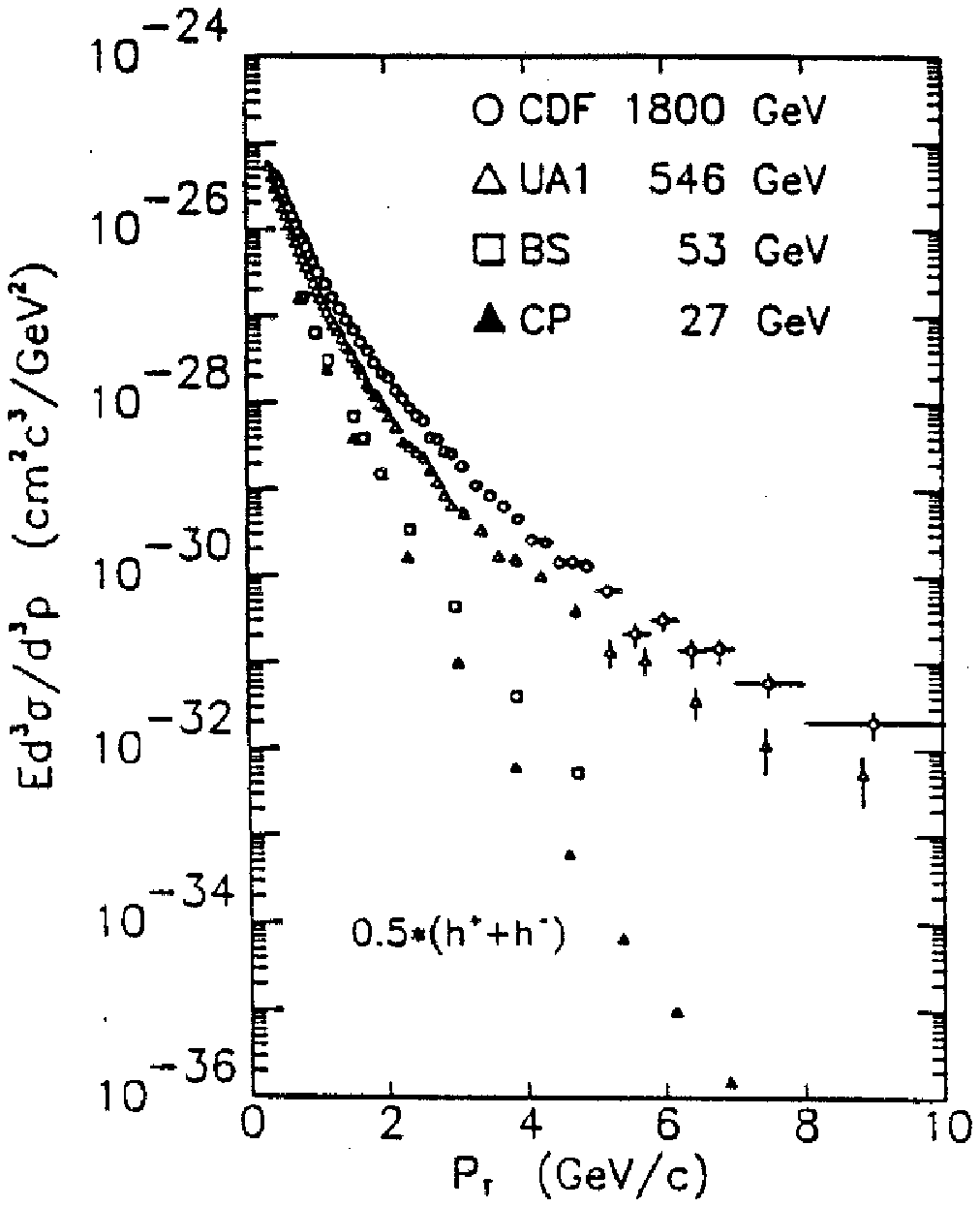}&
\includegraphics[width=0.44\linewidth]{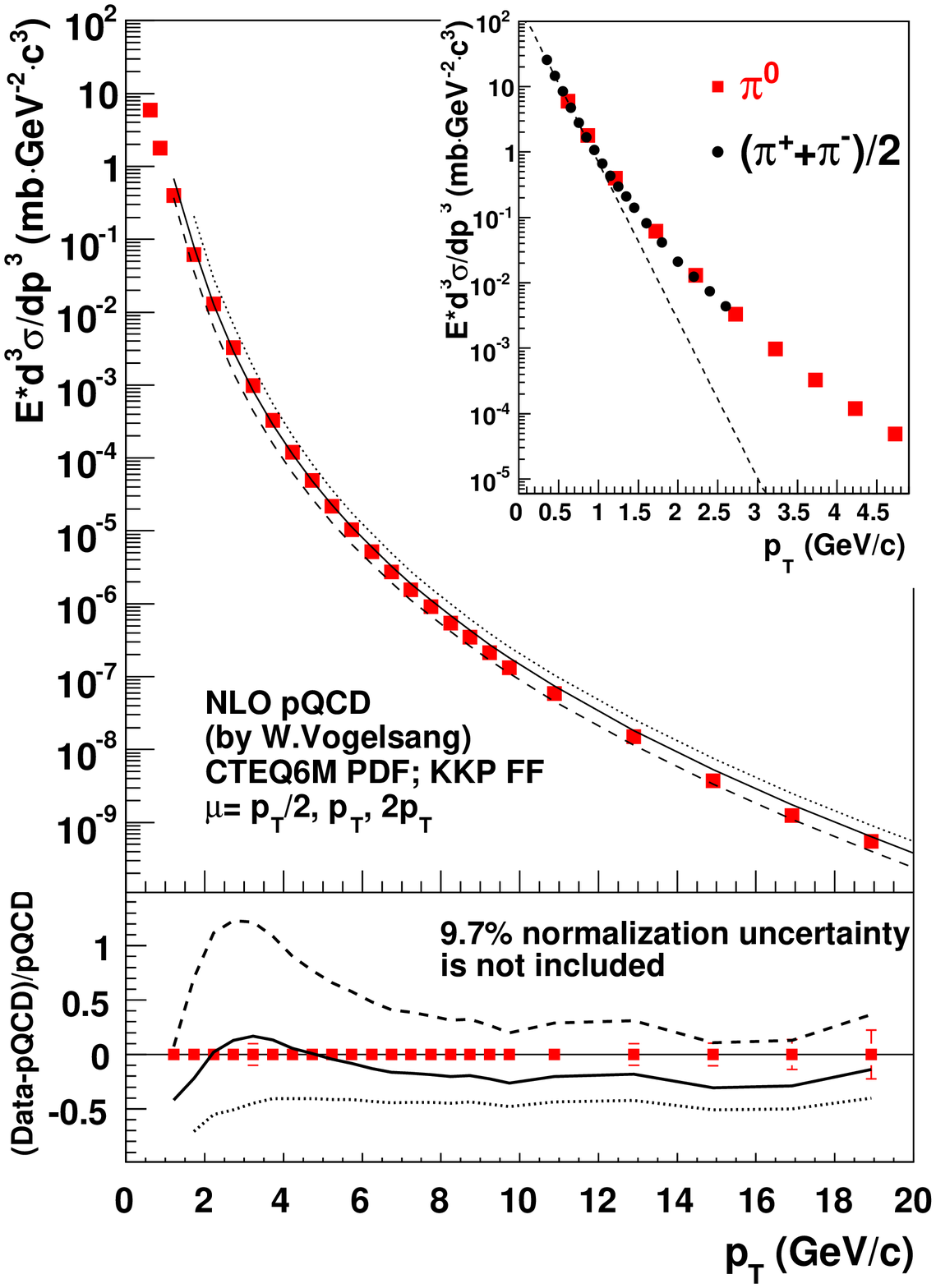}
\end{tabular}
\end{center}\vspace*{-0.25in}
\caption[]{(left) $E {d^3\sigma}/{d^3p}$ for $h^{\pm}$ at mid-rapidity as a function of $p_T$ for several values of $\sqrt{s}$ in p-p collisions~\cite{CDFcompilation}. (right) PHENIX measurement of $\pi^0$ in p-p collisions at $\sqrt{s}=200$ GeV~\cite{ppg063}.  }
\label{fig:thenandnow}
\end{figure}

  Hard-scattering in p-p collisions was discovered at the CERN-ISR in 1972, at c.m. energies $\sqrt{s}=23.5-62.4$ GeV, by the observation of an unexpectedly large yield of particles with large transverse momentum ($p_T$)~\cite{egseeMJTNPA05}. The exponential behavior $e^{-6p_T}$ at low $p_T$ breaks to a power-law tail which varies systematically with the $\sqrt{s}$ of the collision (Fig.~\ref{fig:thenandnow}-(left)). 
Figure~\ref{fig:thenandnow}-(right)~\cite{ppg063} shows the relative composition of the ``hard'' and ``soft'' components of the $p_T$ spectrum in p-p collisions at $\sqrt{s}=200$ GeV. The soft, thermal-like, exponential spectrum, which has changed only slightly to $e^{-5.6 p_T}$, dominates particle production for $p_T\leq 1.5$ GeV/c ($\sim 99.7$\% of the soft particles), while hard scattering predominates for $p_T\geq 2.0$ GeV/c. 

	Measurements of inclusive single or pairs of hadrons at the CERN-ISR were used to establish that high transverse momentum particles in p-p collisions are produced from states with two roughly back-to-back jets which are  the result of scattering of constituents of the nucleons as described by Quantum Chromodynamics (QCD), which was developed during the course of those measurements~\cite{egseeMJTNPA05}. These techniques have been used extensively and further developed at RHIC since they are the only practical method to study hard-scattering and jet phenomena in A+A central collisions due to the large multiplicity, roughly $A$ times that of p-p collisions.  
	
\begin{figure}[ht]
\begin{center}
\includegraphics[width=0.57\linewidth,angle=-90]{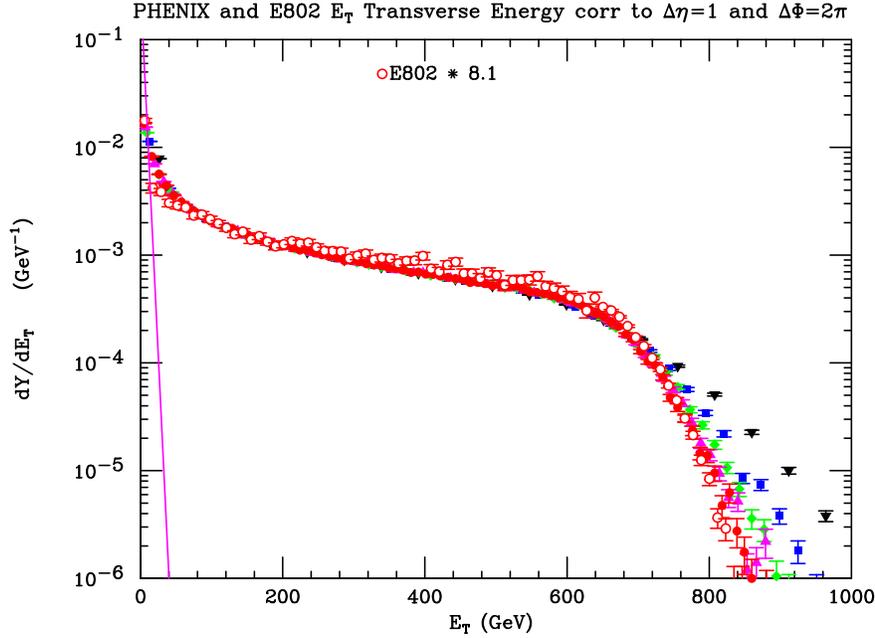} 
\end{center}
\caption[]{$E_T$ distributions in Au+Au collisions: PHENIX at $\sqrt{s_{NN}}=200$ GeV~\cite{PXET} compared to E802 at $\sqrt{s_{NN}}\approx 5$ GeV~\cite{E802ET} scaled up in $E_T$ by 8.1. The quantity plotted is the normalized yield, $dY/dE_T$  (GeV$^{-1}$), which integrates to 1, i.e $\int dE_T\;dY/dE_T=1$. Different filled symbols represent PHENIX measurements in solid angles $\Delta\eta=0.76$, $\Delta\phi=m\times \pi/8$ at mid rapidity with $m=1,...5$. Solid line is an estimate of $E_T$ distribution in p-p collisions at $\sqrt{s}=200$ GeV for $m=5$. }
\label{fig:E802PXpp} 
\end{figure}

	A quantity related to multiplicity, but more relevant for the study of jets, is the distribution of transverse energy, $E_T=\sum_i E_i \sin \theta_i$, where the sum is taken over all particles emitted on an event into a fixed but large solid angle, typically measured in a calorimeter, and then corrected to the solid angle $\Delta\eta=1$, $\Delta\phi=2\pi$~\cite{egseeMJT89-04}.  This is shown in Fig.~\ref{fig:E802PXpp} for Au+Au collisions at AGS (nucleon-nucleon c.m. energy, $\sqrt{s_{NN}}\sim 5.0$ GeV) and RHIC ($\sqrt{s_{NN}}=200$ GeV), 
together with an estimate at RHIC for p-p collisions. The fact that the scaled AGS and RHIC $E_T$ distributions lie one on top of each other shows that the shape of $E_T$ distributions is largely dominated by the nuclear-geometry. Above the upper 0.5\%-ile of the distribution ($E_T\approx 700$ GeV on Fig.~\ref{fig:E802PXpp}),  measurements in smaller solid angles show increasingly larger fluctuations  represented by  flatter slopes above the knee of the distribution. 
	
\begin{figure}[!thb]
\begin{center}
\begin{tabular}{cc}

\includegraphics[width=0.43\linewidth]{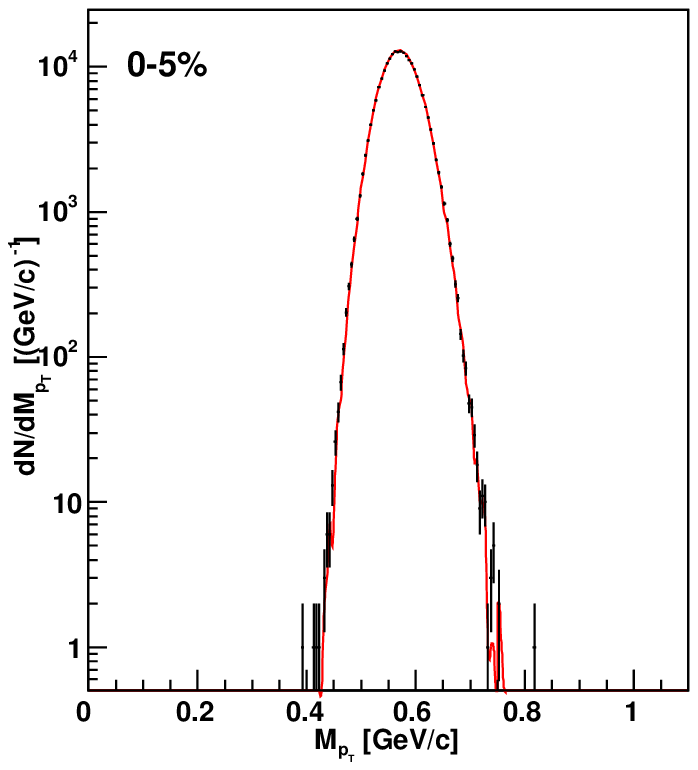}&
\includegraphics[width=0.50\linewidth]{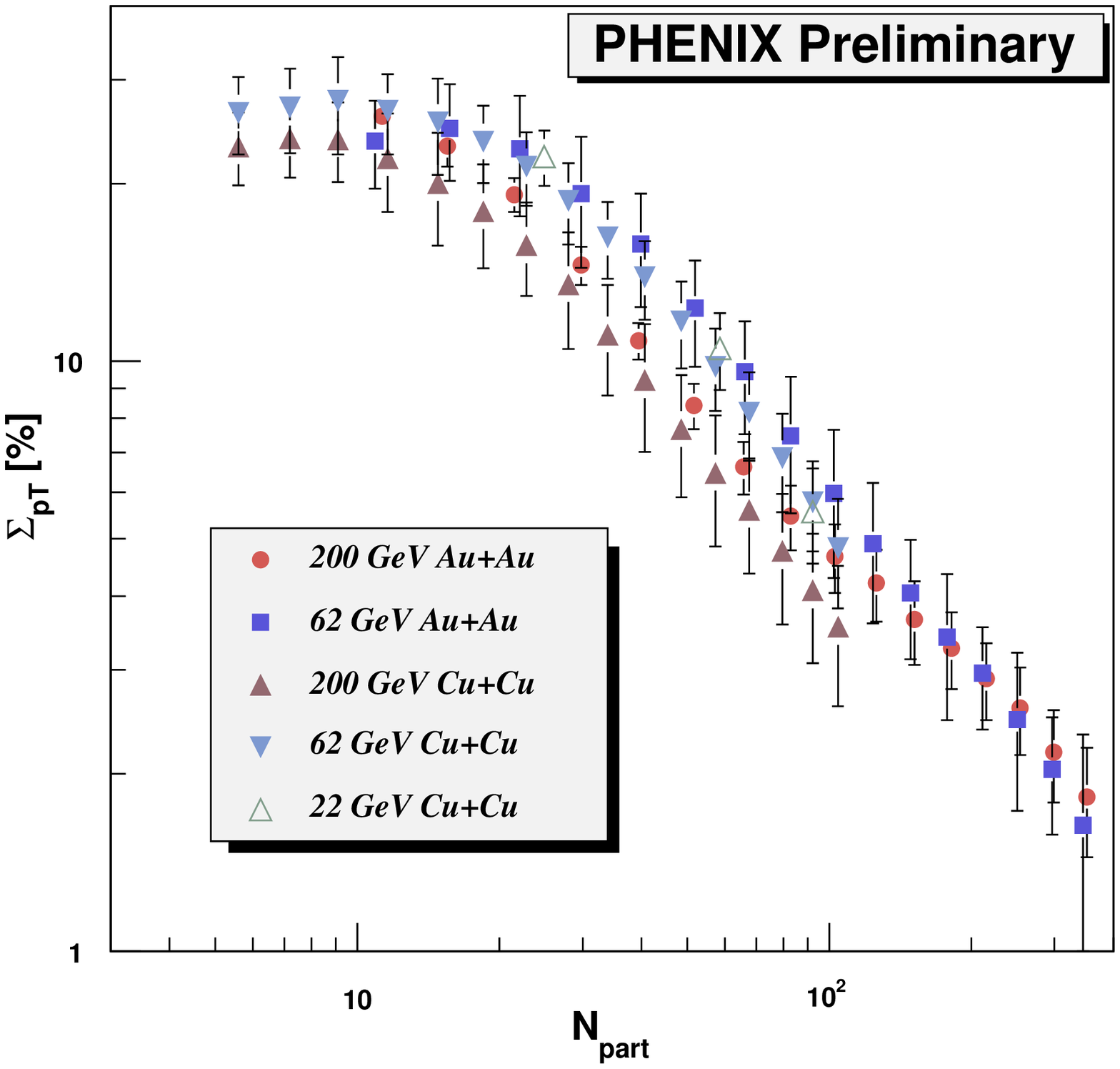}

\end{tabular}
\end{center}\vspace*{-0.25in}
\caption[]{(left) Distribution of $M_{p_T}$ (Eq.~\ref{eq:MpT}) for central Au+Au collisions~\cite{PXPRC66} (data points), random baseline (red line).  (right) Non-random fluctuations~\cite{JTMQM06}, $\Sigma_{p_T}$ for many systems as a function of centrality represented by the number of participants, $N_{\rm part}$.  }
\label{fig:MpT}
\end{figure}

	To test what fraction of this fluctuation is random, the average $p_T$ of particles emitted in an event into a large solid angle,  
	\begin{equation}
	M_{p_T}={1\over n}\sum_{i=1}^{n} p_{T_i}\approx {E_T \over  n} \qquad ,
	\label{eq:MpT}
	\end{equation} 
is measured and compared to a random baseline of mixed events (Fig.~\ref{fig:MpT}-(left)). This indicates that the r.m.s of the random fluctuations is $\sim 5-8$\% of the $\langle E_T\rangle$ for central Au+Au collisions at RHIC and that the non-random fluctuations $\Sigma_{p_T}=\sigma^{NR}_{M_{p_T}}/\mean{M_{p_T}}\simeq 1$\% (Fig.~\ref{fig:MpT}-(right)). The big question for jet analyses in A+A collisions is whether the random or non-random fluctuations are identical in adjacent patches the size of a typical jet cone, $R\equiv\sqrt{(\Delta\eta)^2 +(\Delta\phi)^2}=1$. This point is not merely academic since the energy in a typical jet cone ($R=1$) for central Au+Au collisions at RHIC ($\sqrt{s_{NN}}=200$ GeV) is $\sim 300$ GeV (which is above the maximum jet energy of 100 GeV) while at LHC ($\sqrt{s_{NN}}=5500$ GeV) the energy is 2-5 times larger, 600--1500 GeV for $R=1$, which is at least feasible since it is below the kinematic limit. However, in the author's opinion, looking for jets in heavy ion collisions is like going to the gambling casino: you lose on the average, but you try to win by beating the fluctuations.   

     \section{Hard-scattering formalism}
 
	The overall p-p hard-scattering cross section in ``leading logarithm'' pQCD~\cite{egMJTPoS06,MJTPoS06}   
is the sum over parton reactions $a+b\rightarrow c +d$ 
(e.g. $g+q\rightarrow g+q$) at parton-parton center-of-mass (c.m.) energy $\sqrt{\hat{s}}$:   
\begin{equation}
\frac{d^3\sigma}{dx_1 dx_2 d\cos\theta^*}=
\frac{s d^3\sigma}{d\hat{s} d\hat{y} d\cos\theta^*}=
\frac{1}{s}\sum_{ab} f_a(x_1) f_b(x_2) 
\frac{\pi\alpha_s^2(Q^2)}{2x_1 x_2} \Sigma^{ab}(\cos\theta^*)
\label{eq:QCDabscat}
\end{equation} 
where $f_a(x_1)$, $f_b(x_2)$, are parton distribution functions, 
the differential probabilities for partons
$a$ and $b$ to carry momentum fractions $x_1$ and $x_2$ of their respective 
protons (e.g. $u(x_2)$), and where $\theta^*$ is the scattering angle in the parton-parton c.m. system. 
Equation~\ref{eq:QCDabscat} gives the $p_T$ spectrum of outgoing parton $c$, which then
fragments into a jet of hadrons, including e.g. $\pi^0$.  The fragmentation function
$D^{\pi^0}_{c}(z)$ is the probability for a $\pi^0$ to carry a fraction
$z=p^{\pi^0}/p^{c}$ of the momentum of outgoing parton $c$. Equation~\ref{eq:QCDabscat}
must be summed over all subprocesses leading to a $\pi^0$ in the final state weighted by their respective fragmentation functions. In this formulation, $f_a(x_1)$, $f_b(x_2)$ and $D^{\pi^0}_c (z)$ 
represent the ``long-distance phenomena'' to be determined by experiment;
while the characteristic subprocess angular distributions,
{\bf $\Sigma^{ab}(\cos\theta^*)$} and the coupling constant,
$\alpha_s(Q^2)=\frac{12\pi}{25\, \ln(Q^2/\Lambda^2)}$,
are fundamental predictions of QCD 
for the short-distance, large-$Q^2$, phenomena. One of the original reasons for measuring jets was that it was assumed that the jet cross section would measure the parton cross section (Eq.~\ref{eq:QCDabscat}) without reference to the  fragmentation functions. However, this is not true (see below). 

\subsection{$x_T$ scaling}
 Equation \ref{eq:QCDabscat} leads to a general `$x_T$-scaling' form for the invariant cross
section of high-$p_T$ particle production~\cite{bbk,bbg}: 
\begin{equation}
E \frac{d^3\sigma}{d^3p} = \frac{1}{p_T^{n_{\rm eff}}} F({x_T}) = 
 \frac{1}{\sqrt{s}^{\, n_{\rm eff}}} G({x_T}) \: ,
 \label{eq:bbg}
 \end{equation} 
where $x_T = 2p_T/\sqrt{s}$. 
The cross section has two factors, a function $F({x_T})$ ($G({x_T})$) which `scales',
i.e. depends only on the ratio of momenta, and a dimensioned factor,
${1/p_T^{n_{\rm eff}}}$ ($1/\sqrt{s}^{\, n_{\rm eff}}$),   
where $n_{\rm eff}(x_T,\sqrt{s})$ equals 4  in lowest-order (LO) QCD calculations, analogous to the $1/q^4$ form
of Rutherford scattering in QED. The structure and fragmentation
functions, which scale, since they are functions of the ratios of momenta,  
are all in the
$F(x_T)$ ($G(x_T)$) term. Due to higher-order effects, such as the running of
the coupling constant, $\alpha_s(Q^2)$, the evolution of the
structure and fragmentation functions, and the initial-state
transverse momentum
$k_T$, $n_{\rm eff}(x_T,\sqrt{s})$ is not a constant but is a function of $x_T$, $\sqrt{s}$.
Measured values of ${\,n_{\rm eff} (x_T,\sqrt{s})}$ for $\pi^0$ in p-p 
collisions are between 5 and 8~\cite{egseeMJTNPA05}.

\section{Jets}
\begin{figure}[!thb]
\begin{center}
\begin{tabular}{cc}
\includegraphics[width=0.51\linewidth]{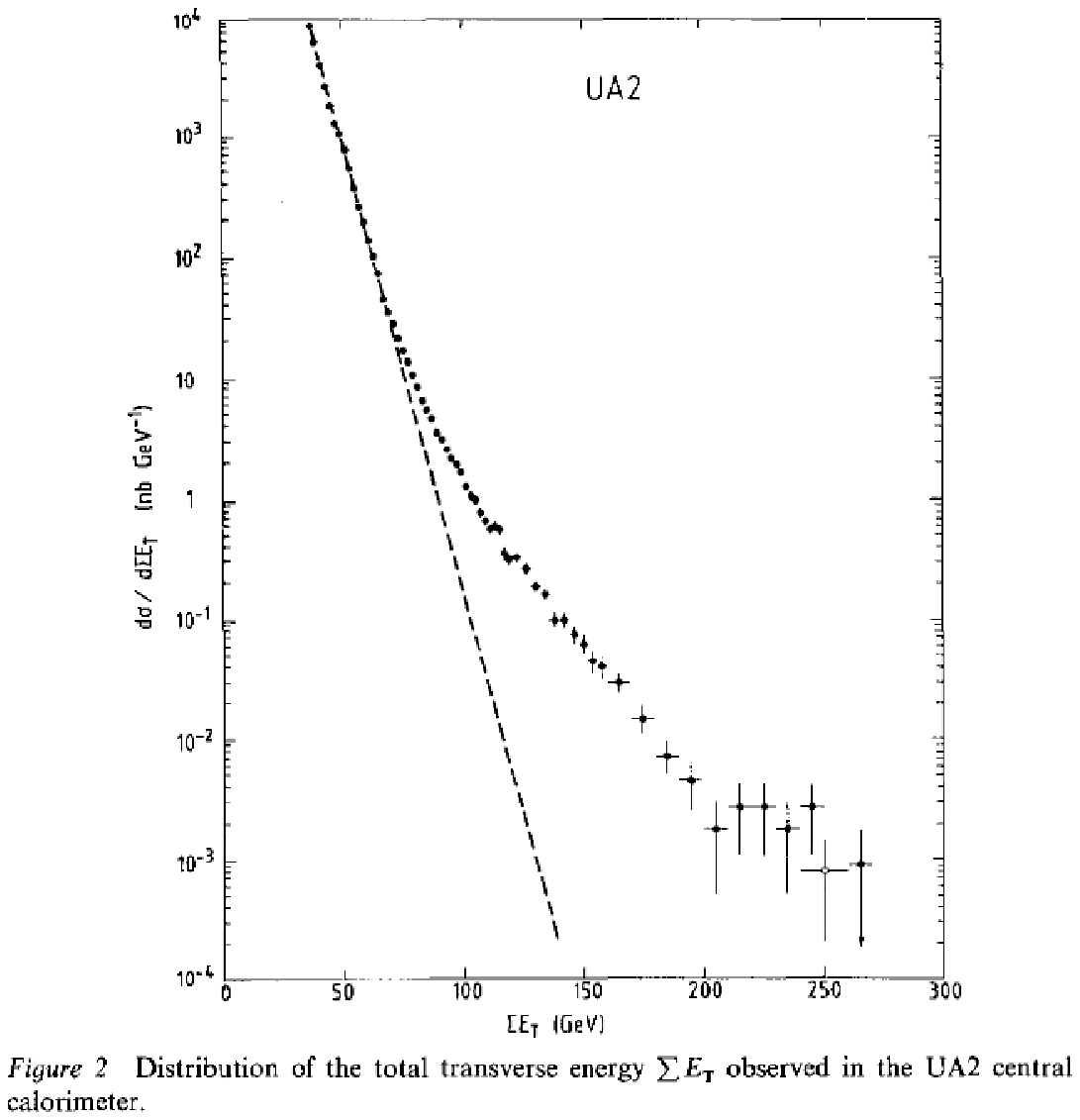}&
\includegraphics[width=0.43\linewidth,angle=+1]{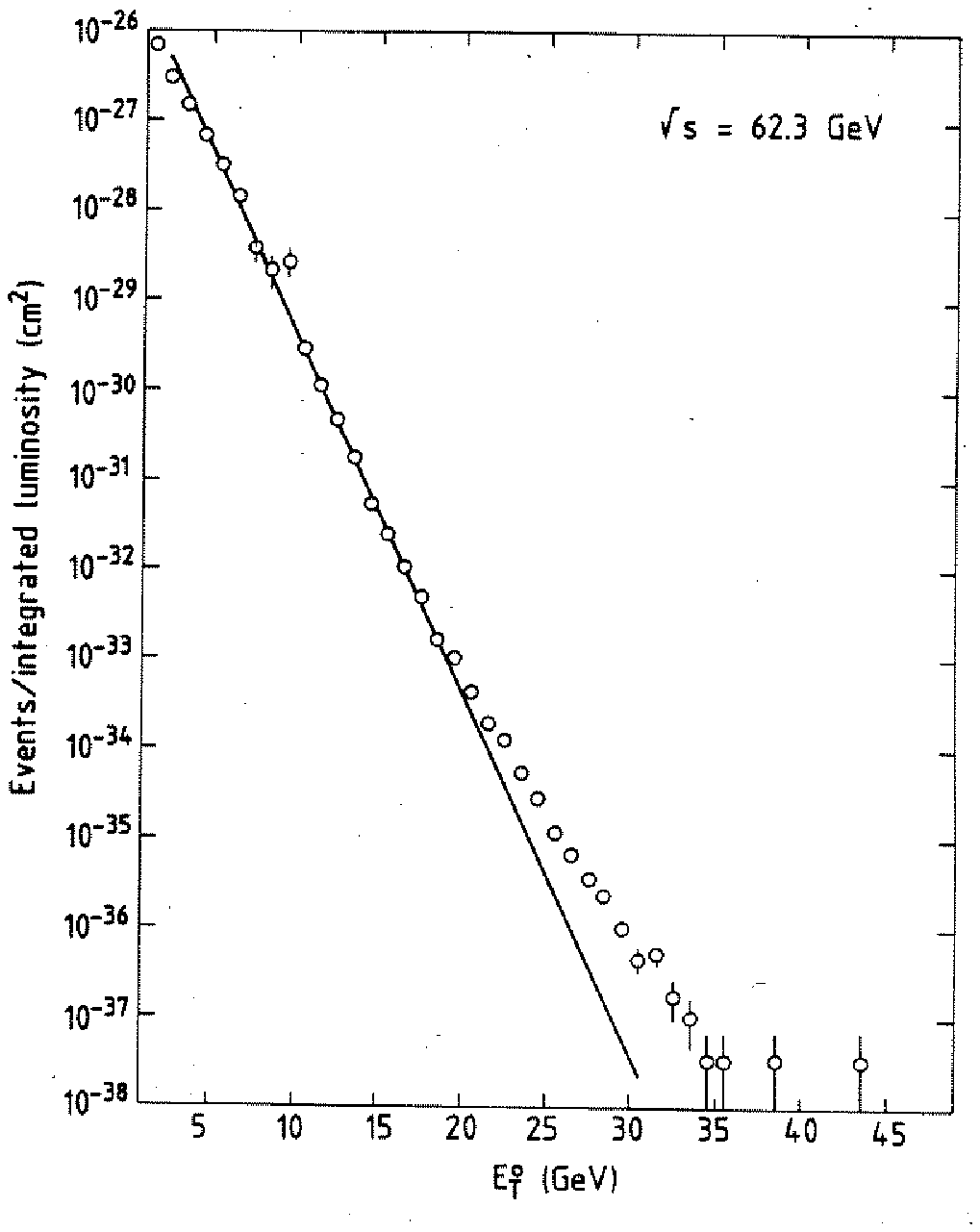}
\end{tabular}
\end{center}\vspace*{-0.25in}
\caption[]{$E_T$ measurements by: (left) UA2 in p-$\bar{\rm p}$ collisions at $\sqrt{s}=540$ GeV~\cite{UA2ET}; (right) CCOR in p-p collisions at $\sqrt{s}=62.3$ GeV~\cite{CCORET}. The lines indicate extrapolation of the nearly exponential spectrum at lower $E_T$.}
\label{fig:ETdists}
\end{figure}

\begin{figure}[ht]
\begin{center}
\includegraphics[width=0.80\linewidth]{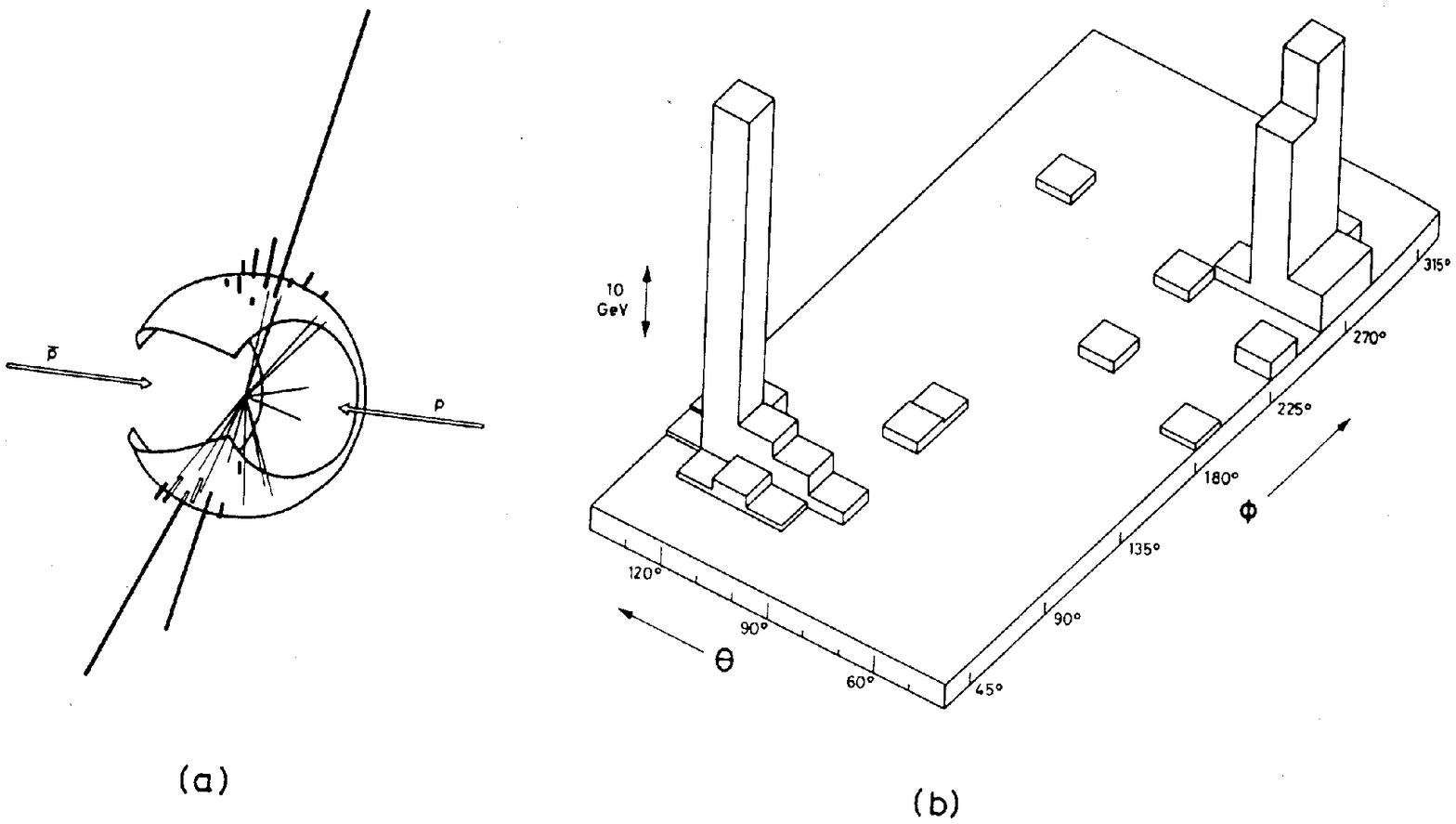} 
\end{center}
\caption[]
{ UA2 jet event from 1982 ICHEP~\cite{Paris82}. a) event shown in geometry of detector. b) ``Lego" plot of energy in calorimeter cell as a function of angular position of cell in polar ($\theta$) and azimuthal ($\Phi$) angle space.
\label{fig:UA2jet} }
\end{figure}

    Hard-scattering is visible by a break in the single particle inclusive spectrum after roughly 3 orders of magnitude in cross section (recall Fig.~\ref{fig:thenandnow}). However, finding jet structure in $E_T$ distributions measured in $4\pi$ calorimeters is much more difficult because the jets are only visible after 5-7 orders of magnitude of the cross section which is even more  dominated by soft physics (see Fig.~\ref{fig:ETdists})~\cite{UA2ET,CCORET}.   
 Due to the fact (which was unknown in the 1970's) that jet structure in $4\pi$ calorimeters at ISR energies or lower is invisible below $\sqrt{\hat{s}}\sim E_T \leq 25$ GeV~\cite{Gordon}, there were many false claims of jet observation in the period 1977-1982. This led to skepticism about jets in hadron collisions, particularly in the USA~\cite{egseeMJT89-04}. A `phase change' in belief-in-jets was produced by one UA2 event 
at the 1982 ICHEP in Paris~\cite{Paris82} (Fig.~\ref{fig:UA2jet}). 
Since that time, jets have become a standard tool of High Energy Physics. 

\begin{figure}[!thb]
\begin{center}
\begin{tabular}{cc}
\includegraphics[width=0.48\linewidth,height=0.40\linewidth]{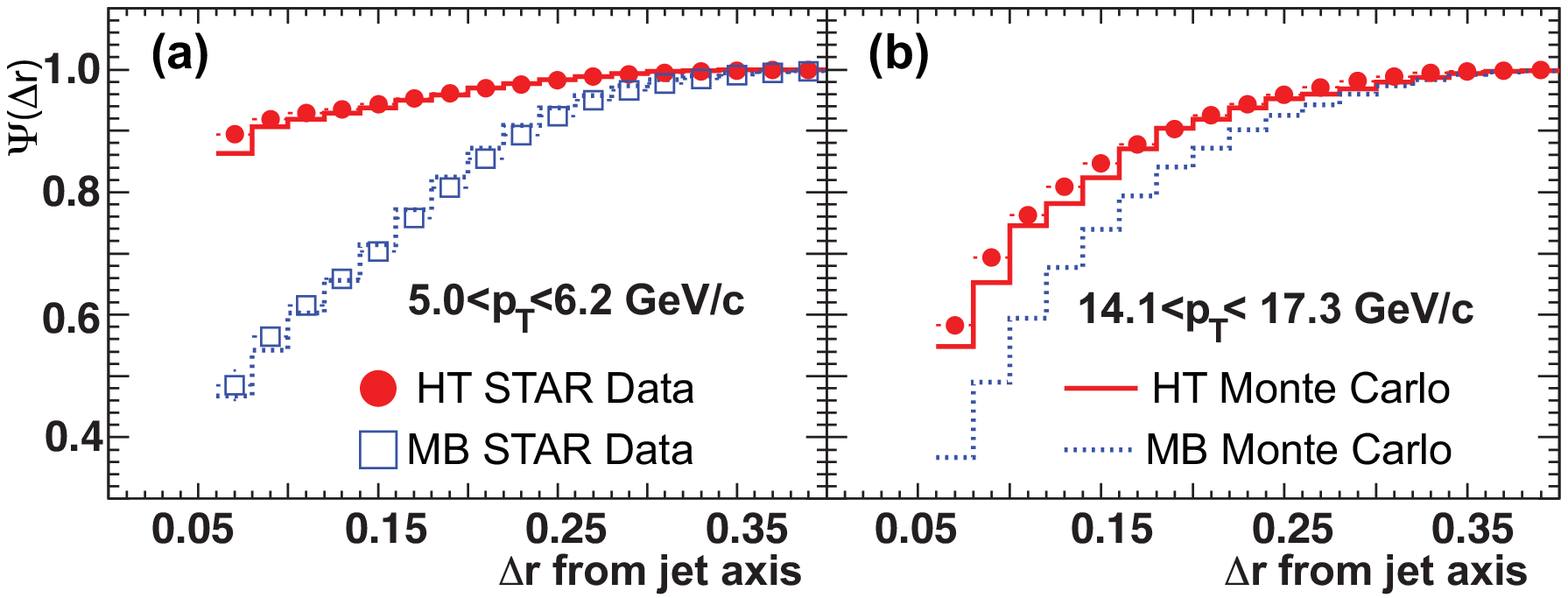}&
\includegraphics[width=0.44\linewidth]{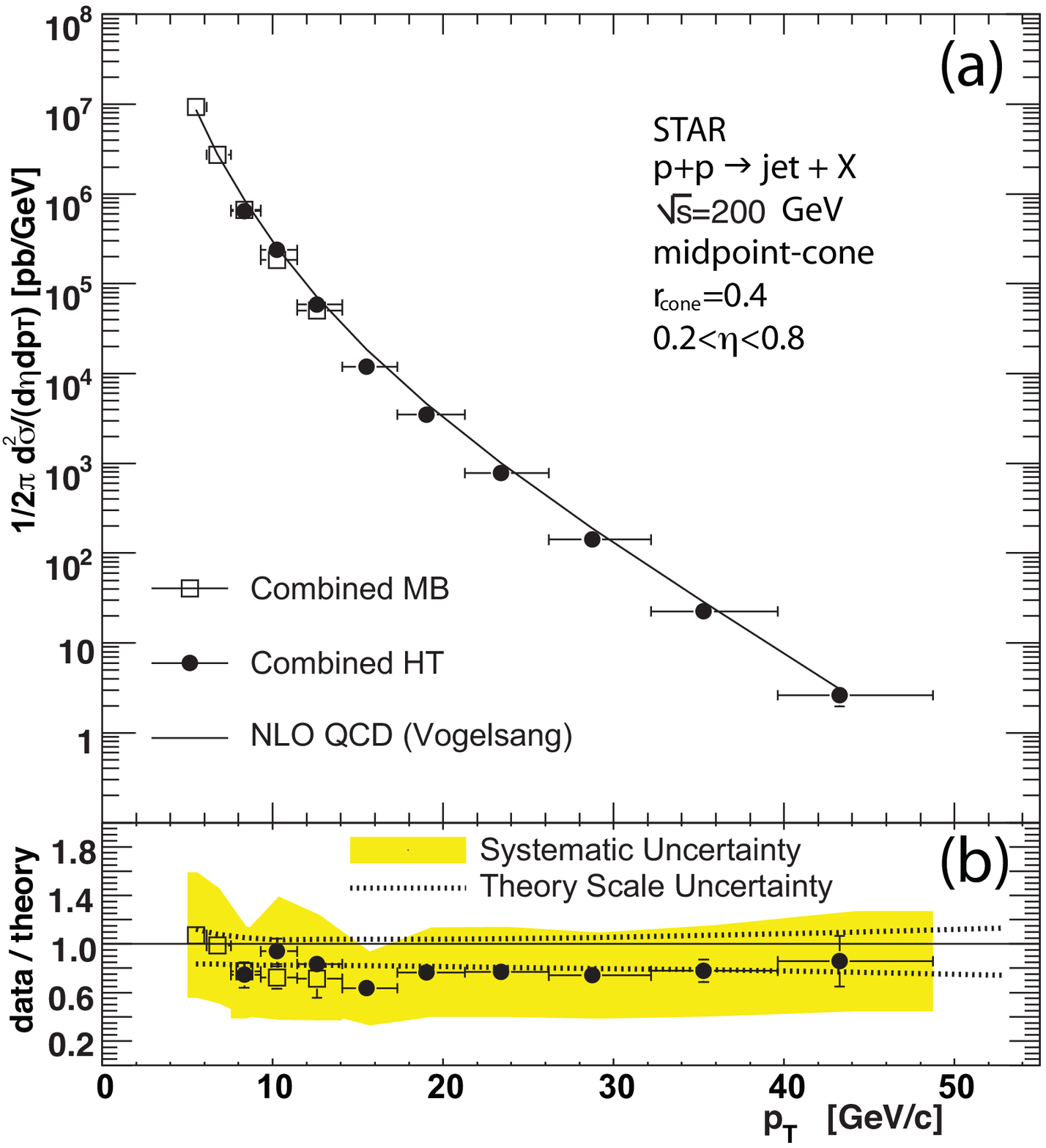}
\end{tabular}
\end{center}\vspace*{-0.25in}
\caption[]{(left) Measured and simulated ``jet profile function'', $\Psi(\Delta r, R, p_T)$, defined as the average fraction of jet $E_T$ inside a coaxial inner cone of radius $\Delta r< r_{\rm cone}$, where $r_{\rm cone}$ is the value of $R\equiv\sqrt{(\Delta\eta)^2 +(\Delta\phi)^2}$ used to define the jet, for minimum bias (MB) events and High Tower (HT) ($R\sim 0.05$) triggers. (right) Non-invariant jet differential cross section vs. jet $p_T$ in the midpoint-cone algorithm for p-p collisions at $\sqrt{s}=200$ GeV~\cite{STARjetPRL97}.  }
\label{fig:STARjet}
\end{figure}

	In fact, this past year, the first measurement of jets in p-p collisions at RHIC (Fig.~\ref{fig:STARjet}) was published~\cite{STARjetPRL97}.
Frankly, I am skeptical of the MB data at low $p_T$ which are too reminiscent of the ``proof by Monte Carlo'' erroneous jet claims of the 1970's~\cite{egMarge,egseeMJT89-04}. However, I think that the HT measurements are reasonable (see below). Note that the systematic error on the cross section is $\pm 49$\%, predominantly due to the 9\% uncertainty in the jet energy ($p_T$?) scale.
\subsection{Internal structure of jets: the ``Humpback'' distribution.}
\begin{figure}[!thb]
\begin{center}
\begin{tabular}{cc}
\includegraphics[width=0.44\linewidth]{figs/cdfMLLA.epsf}&
\includegraphics[width=0.50\linewidth]{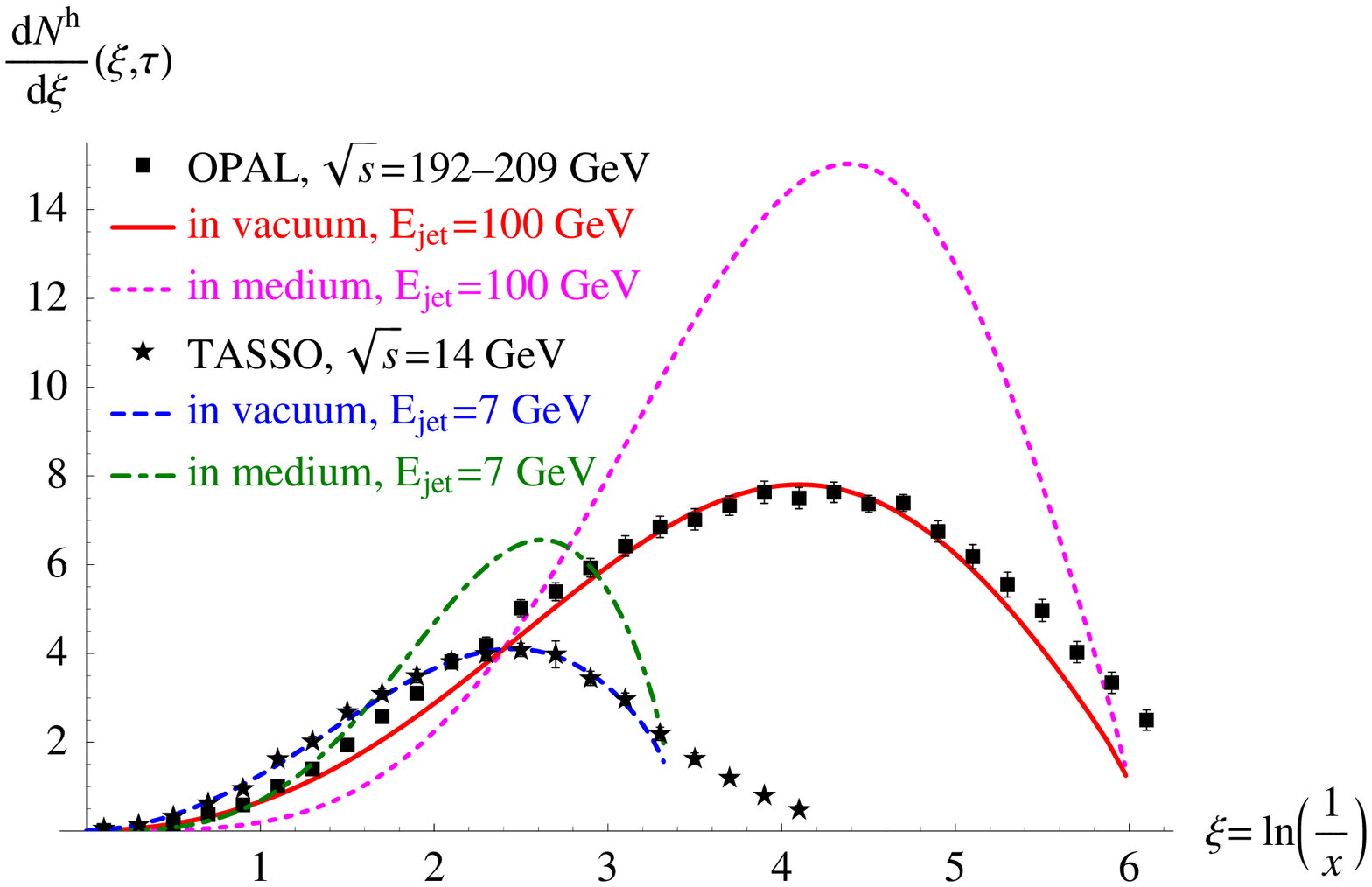}
\end{tabular}
\end{center}\vspace*{-0.25in}
\caption[]{(left) Inclusive momentum distribution of particles in jets of cone size $\theta_c$ in the variable $\xi=\ln(1/z)$~\cite{CDF-PRD68}. (right) Prediction of inclusive distribution of hadrons in jets in vacuum and in a medium~\cite{BorgWiedHump}.  }
\label{fig:cdf}
\end{figure}

   CDF~\cite{CDF-PRD68} has made a beautiful measurement of the evolution with jet energy and cone angle of the jet fragmentation function (Fig.~\ref{fig:cdf}-(left)) as well as the distribution of the momentum of the fragments transverse to the jet axis (which PHENIX~\cite{ppg029} and CCOR call $j_T$ and CDF calls $k_T$).  
The evolution from a parton---which starts at high virtuality $Q^2$ immediately after scattering with $p_T$---to a jet of particles can be described via quark and gluon branching in QCD~\cite{egseeCDF} yielding a ``Humpback'' distribution in the variable $\xi=\ln(1/z)$. Fig.~\ref{fig:cdf}-(right) shows a calculation~\cite{BorgWiedHump}  of the evolution of a parton, both in vacuum and in a medium,  which takes account of quark and gluon branching as well as  scattering (in the medium). A slight shift of the peak from $\xi\sim 4.2$ to $4.6$ ($z=0.015$ to $0.010$) due to the medium is predicted for 100 GeV jets, which corresponds to fragments with 1.5 to 1.0 GeV/c. I expect that these will be exceedingly difficult to dig out from jets in central A+A collisions at LHC. However, the dramatic steepening of the slope of the fragmentation function over the range $\xi < 3.5$ ($z> 0.03$) should be visible and might even correspond to the naive picture of a parton losing energy in the medium and fragmenting in the vacuum with its standard fragmentation function if the plot were also made in the more familiar fragmentation variable $z$.  

	Even if one imagined that jet interactions in the medium in A+A collisions at LHC could be observed as in Fig.~\ref{fig:cdf}, it is important to realize that there are several more complications. For instance, the high rate for jets at LHC may not be all good news. NLO, NNLO, and still higher order effects may cause lots of multi-jets instead of di-jets. Even di-jets may be very complicated because they are produced by all 2-to-2 subprocesses such as $gg\rightarrow gg$, $qg\rightarrow qg$, $gg\rightarrow u \bar{u}$, $gg\rightarrow b\bar{b}$, etc. With all these subprocesses contributing roughly equally at medium $p_T\sim 60$ GeV/c (since QCD is coupled to color not flavor) the jet structure might not be simple even in p-p collisions. 
\subsubsection{Make sure to read the fine print.} 
\begin{figure}[!thb]
\begin{center}
\begin{tabular}{cc}

\includegraphics[width=0.48\linewidth]{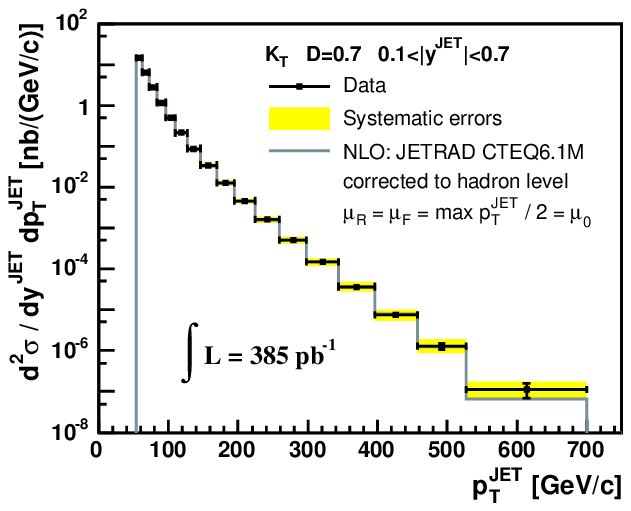}&
\includegraphics[width=0.48\linewidth]{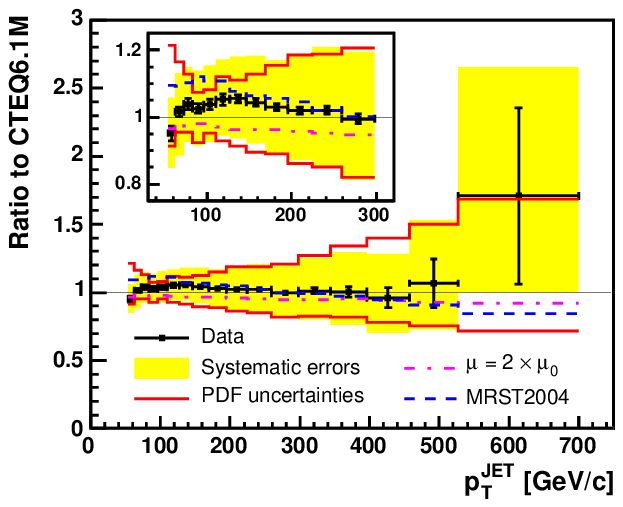}

\end{tabular}
\end{center}\vspace*{-0.25in}
\caption[]{(left) Measured jet cross section compared to theory~\cite{CDFPRL2006} and (right) divided by theory.   }
\label{fig:CDF06}
\end{figure}

Although jet measurements give beautiful results, large sections of the publications read more like legal contracts between experimentalists and theorists  than like scientific papers. From Ref.~\cite{CDF-PRD68}: ``The energy of a jet is defined as the sum of the energies of the towers belonging to the corresponding cluster. Corrections are applied to compensate for the non-linearity and non-uniformity of the energy response of the calorimeter, the energy deposited inside the jet cone from sources other than the parent parton, and the parent parton energy that radiates out of the jet cone. Full details of this procedure can be found in~\cite{Abe92}.'' Reference ~\cite{Abe92} continues with many ($\geq5$) pages of description of corrections and uncertainties. Nevertheless, I must admit that jet measurements of QCD in p-p collisions are now standard after a $\sim 30$ year learning curve (see Fig.~\ref{fig:CDF06})~\cite{CDFPRL2006}.
However, even in describing this beautiful result, there is a qualification~\cite{CDFPRL2006}:``The measured cross section is in agreement with NLO pQCD predictions {\em after the necessary parton-to-hadron corrections are taken into account}.'' In plain English, this means that the jet cross section is not simply the parton cross section as originally hoped. 

\subsection{My favorite jet measurement.} 
\begin{figure}[!thb]
\begin{center}
\begin{tabular}{cc}
\includegraphics[width=0.48\linewidth]{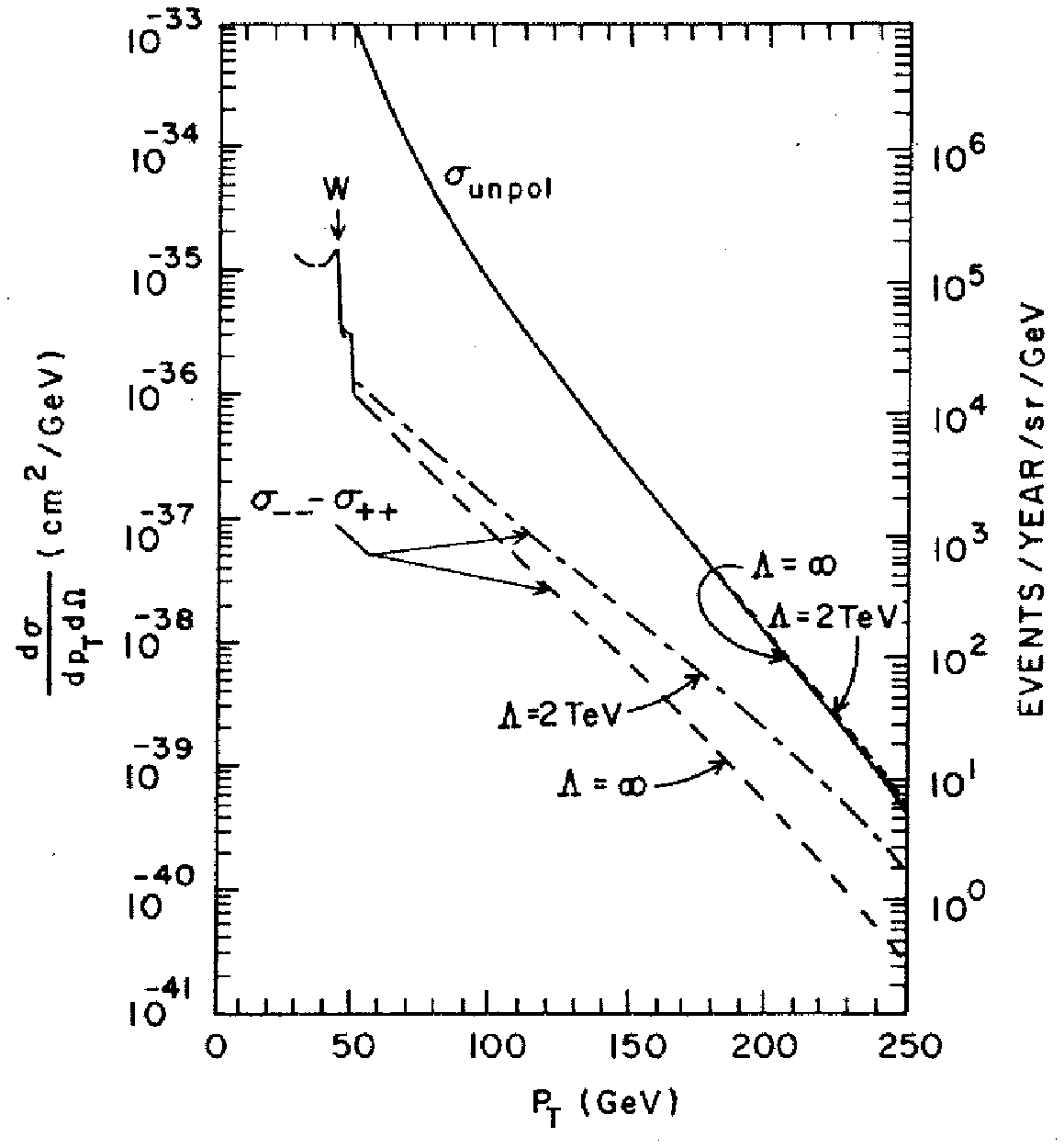}&
\includegraphics[width=0.48\linewidth]{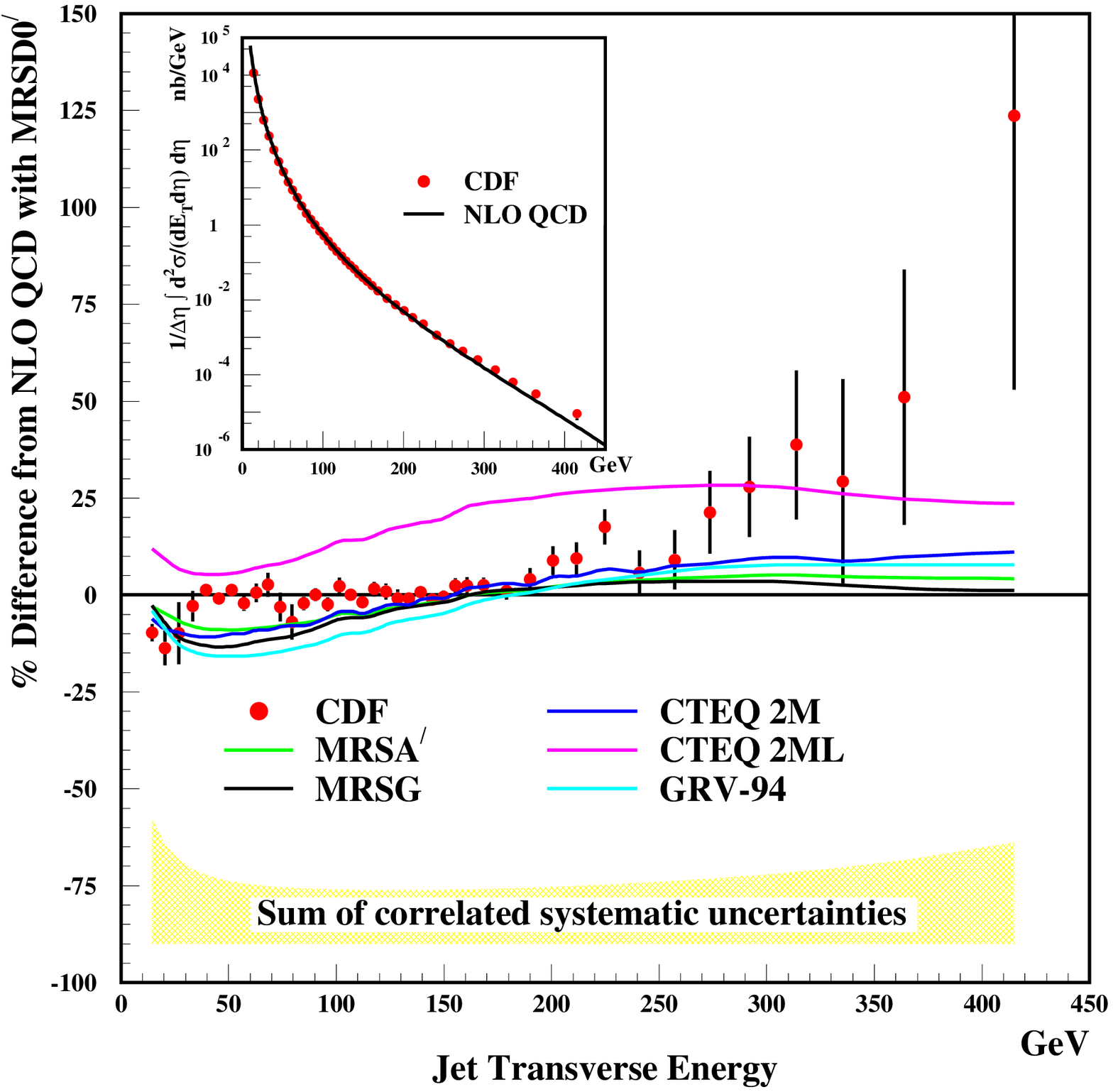}
\end{tabular}
\end{center}\vspace*{-0.25in}
\caption[]{(left) Prediction~\cite{FPMJT} from 1983 for the effect of quark substructure on inclusive jet cross section with and without Parity Violation capability. (right) CDF inclusive jet cross section circa 1996 and ratio to NLO QCD.~\cite{CDFLambda}  }
\label{fig:jetpv}
\end{figure}

   Just to show that I am not totally against jets, one of my favorite measurements in High Energy Physics, and one that I proposed to perform at RHIC~\cite{FPMJT}, is actually a jet measurement (Fig.~\ref{fig:jetpv}-(left)).
One possible explanation of the several generations of quarks and leptons is
that they are composites of more fundamental constituents~\cite{ELP}. Such models of quark substructure generally violate parity since the scale of compositeness is above the mass of the $W^{\pm}$. The parity-violating asymmetry ($\sigma_{--} - \sigma_{++}$) then provides
direct and much more quantative tests for substructure than other methods since the increase in the jet cross section at large $p_T$ due to the substructure is small, roughly the thickness of the line used to draw the figure (see Fig.~\ref{fig:jetpv}-(left)). Interestingly, CDF observed such a deviation from QCD in 1996~\cite{CDFLambda} (Fig.~\ref{fig:jetpv}-(right)-inset). If Fig.~\ref{fig:jetpv}-(right) represented ``\% Parity Violation'' rather than ``\% Difference from NLO QCD'', it would be an unambiguous proof of new physics. However, the CDF deviation was ``explained'' by invoking an ``improved'' Gluon structure function at large $x$~\cite{CTEQ}. This explanation is not credible unless the Gluon structure function can be verified  by other means such as by measurements of direct-photon production with $x>0.25$. 

	For the record, the reason that parity violation measurements to search for quark compositeness at RHIC must be done with $\sim 100$ GeV jets rather than $\pi^0$'s is because the rate for $\sim 100$ GeV $\pi^0$ is smaller by $\sim 2$ orders of magnitude (see below).

\section{Hard scattering via single particle inclusive and few particle correlation measurements}
Single particle inclusive and two or three particle correlation measurements of hard-scattering have provided a wealth of discoveries at RHIC. Due to the steeply falling power-law ($\hat{p}_{T_t}^{-n}$) invariant transverse momentum spectrum of the scattered parton, the inclusive single particle (e.g. $\pi^0$) $p_{T_t}$ spectrum from jet fragmentation is dominated by fragments with large $z_t$, where $z_t=p_{T_t}/\hat{p}_{T_t}$ is the fragmentation variable, and exponential fragmentation $D^{\pi^0}_q(z)\sim e^{-bz}$ is assumed. This gives rise to several effects which allow precision measurements of hard scattering to be made 
using single inclusive particle spectra and two particle correlations~\cite{egMJTPoS06}.

   The most famous properties in this regard are the ``leading-particle effect''
also known as `trigger bias', although  it has nothing to do with the use of a hardware trigger, and the Bjorken ``parent-child relationship''. The Bjorken `parent-child' effect is that the $\pi^0$ spectrum from hard scattering of a parton $q$ has the same power $n$ as the parton spectrum, with a ratio at a given $p_{T_t}=\hat{p}_{T_t}$ of~\cite{MJTPoS06}: 
\begin{equation} 
\left . {\pi^0 \over q }\right|_{\pi^0} (p_{T_t})  \approx {\Gamma(n-1) \over b^{n-3}} \qquad , 
\label{eq:pioverjet}
\end{equation} 
where the normalization gives the total parton cross section assuming equal fragmentation to $\pi^0$, $\pi^+$, $\pi^-$.
Similarly, although the $\mean{z}$ of a fragment $\pi^0$ from a jet, given a parton of $\hat{p}_{T_t}$, is $\mean{z}|_q=1/b$, the $\mean{z_t}$ of a $\pi^0$ in a spectrum of fragments from a power law distribution of partons is: 
\begin{equation}
 \mean{z_t(p_{T_t})}|_{\pi^0}\approx {{n-1} \over b}\qquad ,
 \label{eq:ans1_meanz_inc}
\end{equation}
 a factor of $n-1$ times larger than the the unconditional $\mean{z}$ of fragmentation and independent of $p_{T_t}$.  This is the ``leading-particle effect'', or ``trigger-bias'', which is larger for steeper parton spectra (larger $n$, lower $\sqrt{s}$.).  
  
	The prevailing opinion from the 1970's until last year was that although the inclusive single particle (e.g. $\pi^0$) spectrum from jet fragmentation is dominated by trigger fragments with large $\mean{z_t}\sim 0.6-0.8$, the away-jets should be unbiased and would measure the fragmentation function, once the correction is made for $\mean{z_t}$ and the fact that the jets don't exactly balance $p_T$ due to the $k_T$ smearing effect~\cite{FFF}. Last year, it was found by explicit calculation that this is not true~\cite{ppg029,egMJTPoS06}. The $p_{T_a}$ spectrum of fragments from the away-side parton with $\hat{p}_{T_a}$, given a trigger particle with $p_{T_t}$ (from a trigger-side parton with $\hat{p}_{T_t}$), is not sensitive to the shape of the fragmentation function ($b$), but measures the ratio of $\hat{p}_{T_a}$ of the away-parton to $\hat{p}_{T_t}$ of the trigger-parton and depends only on the same power $n$ as the invariant single particle spectrum:  
		     \begin{equation}
\left.{dP_{p_{T_a}} \over dx_E}\right|_{p_{T_t}}\approx {\mean{m}(n-1)}{1\over\hat{x}_h} {1\over
{(1+ {x_E \over{\hat{x}_h}})^{n}}} \, \qquad . 
\label{eq:condxe2}
\end{equation}
This equation gives a simple relationship between the ratio $x_E\approx p_{T_a}/p_{T_t}$ of the transverse momenta of the away-side particle to the trigger particle, and the ratio of the transverse momenta of the away-jet to the trigger-jet, $\hat{x}_{h}=\hat{p}_{T_a}/\hat{p}_{T_t}$. The only dependence on the fragmentation function is in the mean multiplicity $\mean{m}$ of jet fragments. Note that Eq.~\ref{eq:condxe2} is a Hagedorn function $1/(1+x_E/\hat{x}_h)^n$ and exhibits ``$x_E$-scaling'' in the variable $x_E/\hat{x}_h$. 
\subsection{Calculation of the parton cross section from the $\pi^0$ spectrum}
	 Since the STAR jet measurement (Fig.~\ref{fig:STARjet}) is based primarily on the High Tower neutral cluster trigger, and since a $\pi^0$ is a very well defined neutral cluster, and since the $\pi^0$ invariant $p_T$ spectrum for $p_T\geq 3$ GeV/c is a pure power law with $n=8.10\pm 0.05$~\cite{ppg054}, it occurred to me that the Bjorken parent-child relation, Eq.~\ref{eq:pioverjet}, could be used to calculate the parton cross section from the $\pi^0$ spectrum of Fig.~\ref{fig:thenandnow}-(right)~\cite{ppg063}. The only additional inputs required are the fragmentation functions for quark jets, $D_q(z)\approx \exp -(8.2z)$, and gluon jets, $D_q(z)\approx \exp -(11.4z)$, which were obtained from LEP data in Ref.~\cite{ppg029}, and the relative contribution of quarks and gluons in the scattered parton spectrum (see Table~\ref{tab:piqratios}). 
	 
\begin{figure}[!thb]
\begin{center}
\begin{tabular}{c}
\includegraphics[width=0.66\linewidth]{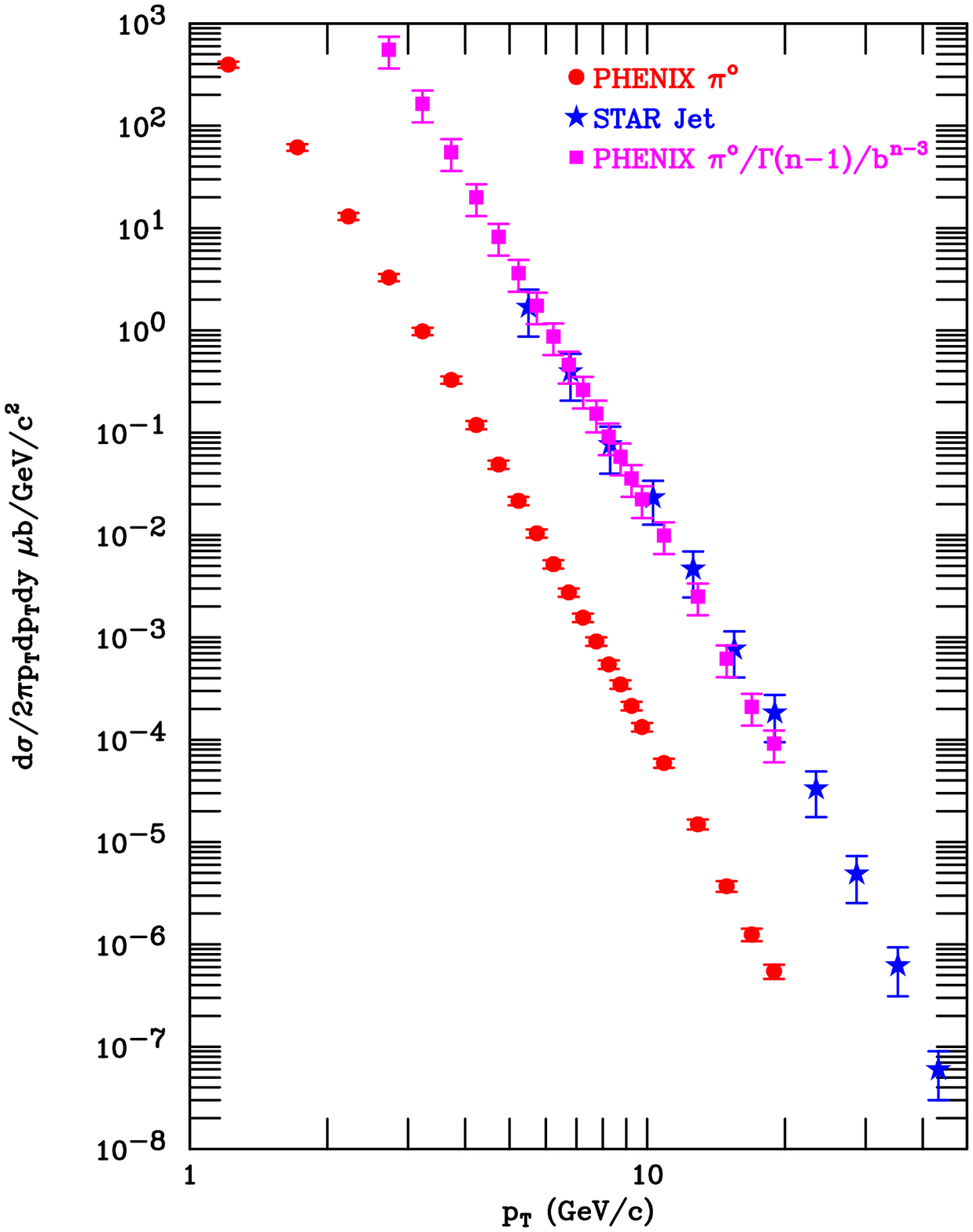}\\
\hspace*{0.30in}\includegraphics[height=0.66\linewidth,angle=90]{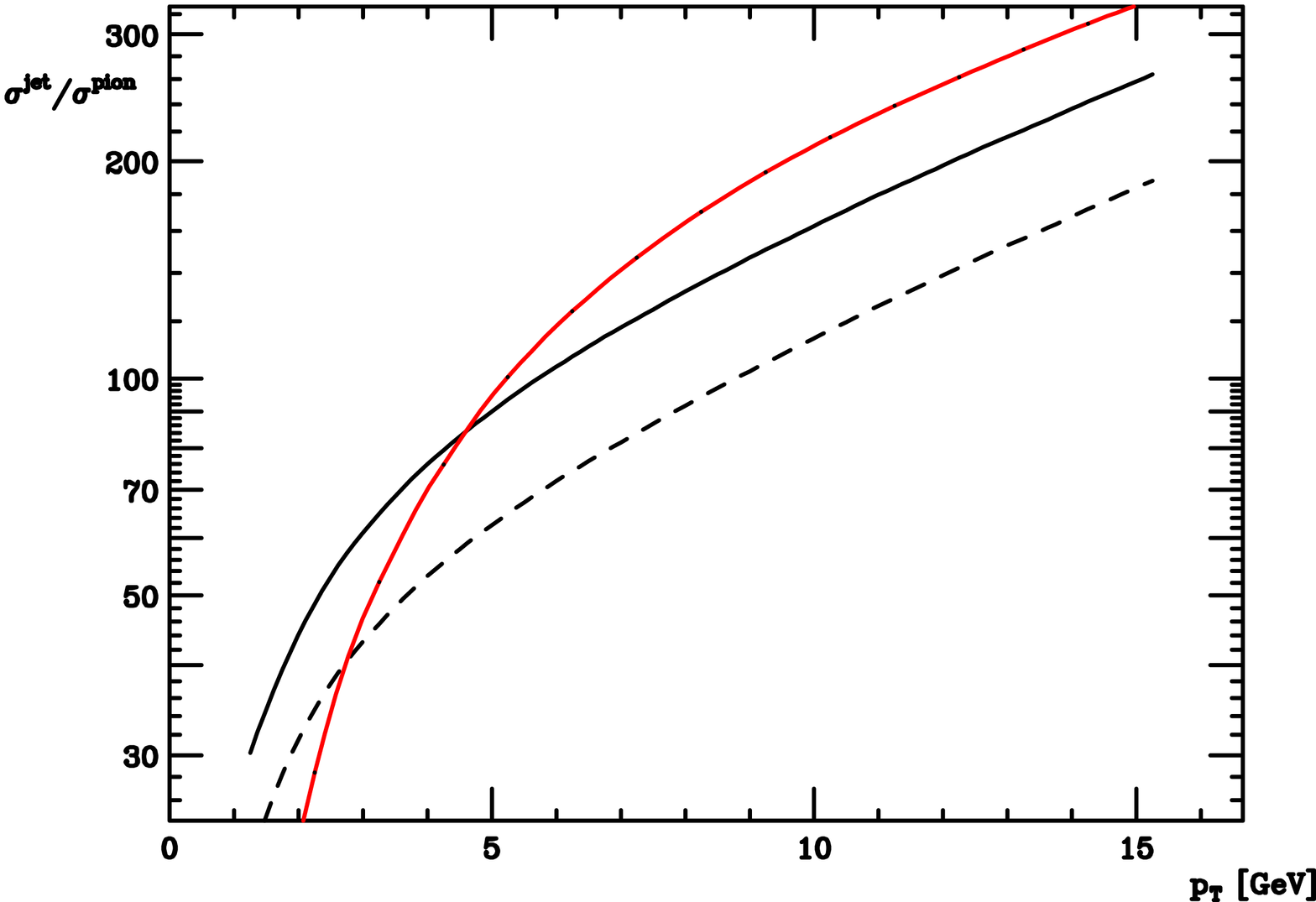}
\end{tabular}
\end{center}\vspace*{-0.25in}
\caption[]{(top) PHENIX $\pi^0$ invariant cross section in p-p collisions $\sqrt{s}=200$ GeV from Fig.~\ref{fig:thenandnow}-(right) (filled circles) divided by Eq.~\ref{eq:pioverjet} for  ${1\over 2}q$ and ${1\over 2}g$ parton compositon, $f_g=0.5$ (Table~\ref{tab:piqratios}) (filled squares). Error bars correspond to $f_g=0.75$ and $f_g=0.25$. STAR data from Fig.~\ref{fig:STARjet} (stars). (bottom) Plot by Werner Vogelsang of jet/pion NLO cross sections ($\sigma^{\rm jet}/\sigma^{\rm pion}$) for all jets (red), quarks-only   (black) and quarks-only LO (black-dashed)~\cite{thanksWV}.  }\vspace*{-0.25in}
\label{fig:PHENIXjet}
\end{figure}

	 	 \begin{table}[h]
\begin{center}
\begin{tabular}{lccccc}
\hline
\multicolumn{1}{c}{Gluon Fraction $f_g=1-f_q$} & 1.0 & 0.75  &  0.50 & 0.25 & 0.0 \\ \hline
$q/\pi^0|_{\pi^0}=1/ ( \pi^0/q|_{\pi^0})$ & 283  & 225  &  168    & 110 & 52.7\\
\hline
\end{tabular}
\caption[]{\label{tab:piqratios} Parton inclusive cross section divided by $\pi^0$ inclusive cross section from Eq.~\ref{eq:pioverjet} as a function of the fraction of gluons and quarks in the parton spectrum.}
\end{center}
\end{table}
The calculated invariant parton cross-section for a composition of 1/2 quarks and 1/2 gluons is shown in Fig.~\ref{fig:PHENIXjet}-(left) and is in   
excellent agreement with the STAR jet measurement~\cite{STARjetPRL97}. However, theoretical calculations~\cite{thanksWV} which agree with both the $\pi^0$ and jet measurements give much different jet/$\pi^0$ and quark-jet/quark-$\pi^0$ cross section ratios---with a completely different trend vs $p_T$---than the simple parton calculations in Table~\ref{tab:piqratios}. This is further illustration that the terms `jet' and `parton' are not equivalent.   
\subsection{Precision measurements of jet suppression in Au+Au collisions}
    Since hard scattering is point-like, with distance scale $1/p_T< 0.1$ fm, the cross section in p+A (B+A) collisions, compared to p-p, should be simply proportional to the relative number of possible point-like encounters, a factor of $A$ ($BA$) for p+A (B+A) minimum bias collisions. For semi-inclusive reactions in centrality class $f$ at impact parameter $b$, the scaling is proportional to $\mean{T_{AB}}|_{f}$, the overlap integral of the nuclear thickness functions averaged over the centrality class. 
\begin{figure}[!thb]
\begin{center}
\begin{tabular}{cc}
\includegraphics[width=0.46\linewidth]{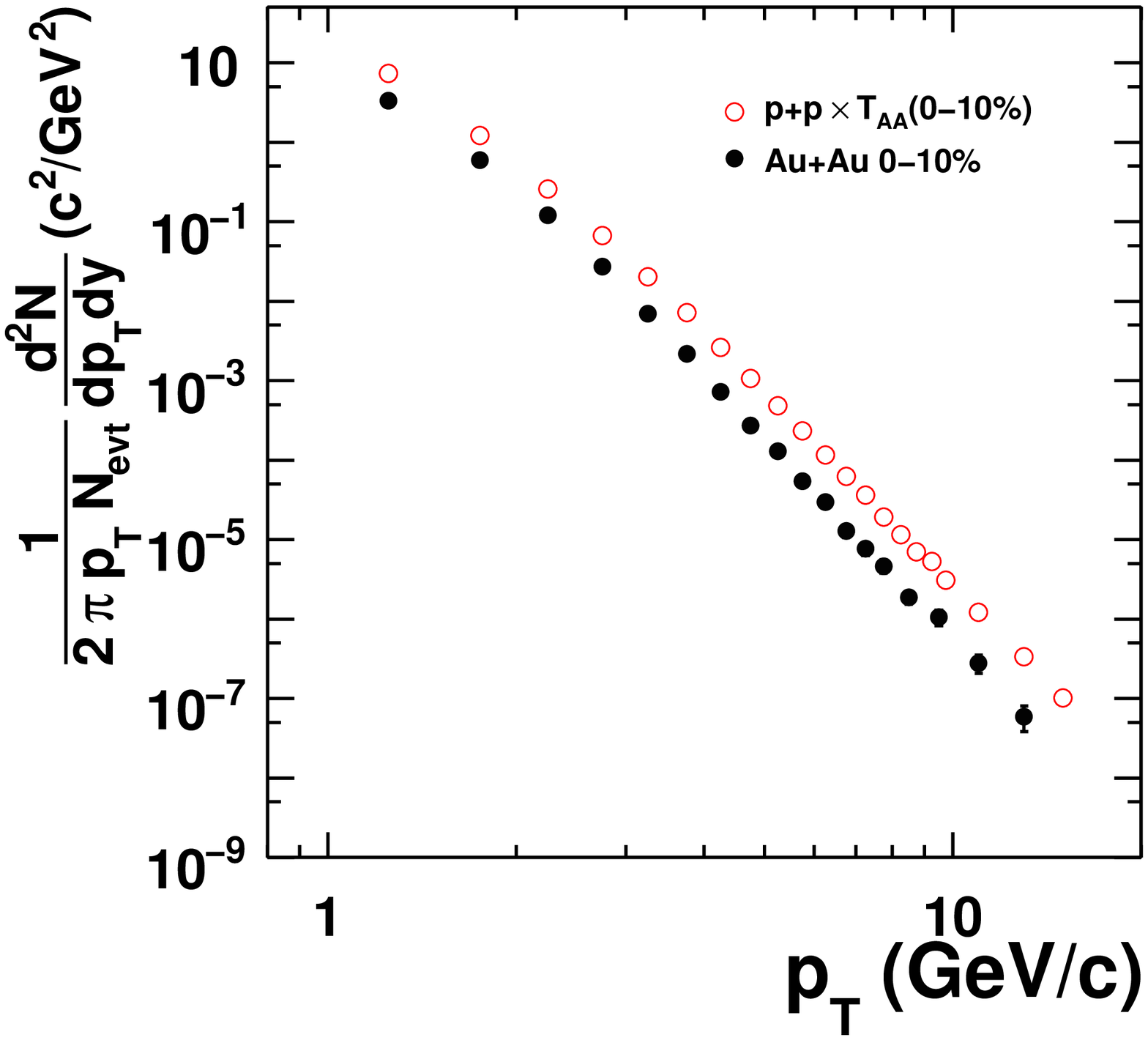}&
\hspace*{-0.02\linewidth}\includegraphics[width=0.54\linewidth]{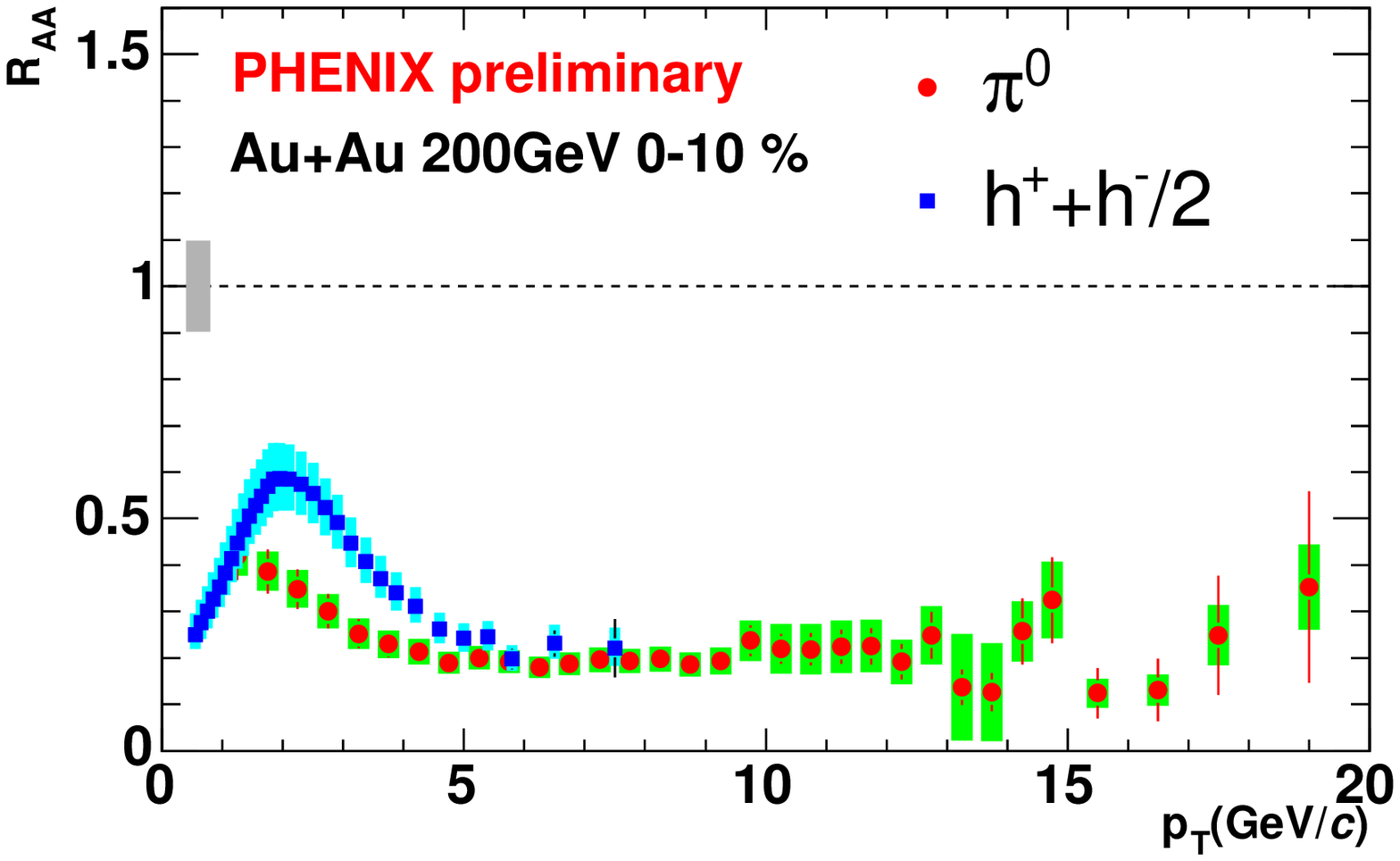}
\end{tabular}
\end{center}\vspace*{-0.25in}
\caption[]{(left)log-log plot~\cite{ppg054} of $\pi^0$ invariant cross section in p-p collisions at $\sqrt{s}=200$ GeV multiplied by $\mean{T_{AA}}$ for Au+Au central collisions (0-10\%) compared to the measured semi-inclusive invariant yield of $\pi^0$. (right) $R_{AA}(p_T)$ for $\pi^0$ and $h^{\pm}$ for Au+Au central (0-10\%) collisions at $\sqrt{s_{NN}}=200$ GeV~\cite{Maya}.  }
\label{fig:RAA}
\end{figure}
    Effects of the nuclear medium, either in the initial or final state, may modify the point-like scaling. This is shown rather dramatically in  Fig.~\ref{fig:RAA}-(left) where the Au+Au data are suppressed relative to the scaled p-p data by a factor of $\sim 4-5$ for $p_T\geq 3$ GeV/c. A quantitative evaluation of the  suppression is made using the ``nuclear modification factor'', $R_{AB}$, the ratio of the measured semi-inclusive yield to the point-like scaled p-p cross section: 
\begin{equation}
R_{AB} = \frac{ d^2 N_{AB}^P / dp_T dy\, N^{inel}_{AB} }{\langle T_{AB} \rangle  \times d^2 \sigma_{pp}^P/dp_T dy}
       \label{eq:RAB}
\end{equation}
where $d^2 N_{AB}^P/dp_T dy\, N^{inel}_{AB}$ is the differential yield of a point-like process $P$ in an $A+B$ collision and $d^2\sigma_{NN}^P/dp_T dy$ is the cross section of $P$ in a p-p collision. For point-like scaling, $R_{AB}\approx 1$. The ratio of point-like scaled measurements in central and  peripheral A+B collisions ($R_{CP}$) is also used to quantify suppression if p-p data are unavailable.  

	While the suppression of $\pi^0$ at a given $p_T$ in Au+Au compared to the scaled p-p spectrum may be imagined as a loss of these particles due to, for instance, the stopping or absorption of a certain fraction of the parent partons in an opaque medium, it is evident from Fig.~\ref{fig:RAA}-(left) that an equally valid quantitative representation can be given by a downshift of the scaled p-p spectrum due to, for instance, the energy loss of the parent partons in the medium---a particle with $p{'}_T$ in the scaled p-p spectrum is shifted in energy by an amount $S(p_T)$ to a measured value $p_T=  p{'}_{T} -S(p_T)$ in the Au+Au spectrum~\cite{ppg054,egMJTPoS06}. 
	The fact that the Au+Au and reference p-p spectra are parallel on Fig.~\ref{fig:RAA}-(left) provides graphical evidence that the fractional $p_T$ shift in the spectrum, $S(p_T)/p_T$, is a constant for $p_T > 3$ GeV/c, which, due to the power law, results in a constant ratio of the $\pi^0$ Au+Au to pp spectra (i.e. $R_{AA}(p_T)=$ constant) as shown in Fig.~\ref{fig:RAA}-(right). 

	    Fig.~\ref{fig:RAA}-(right) also shows that the nuclear modification factors are clearly different for $\pi^0$ and $(h^+ + h^-)/2$ for $p_T < 6$ GeV/c. Is it possible to tell whether one or both of these reactions obey QCD?

\subsection{$x_T$ scaling in A+A collisions as a test of QCD}  
    
\begin{figure}[tbhp]
   \begin{center}
\includegraphics[width=0.8\linewidth]{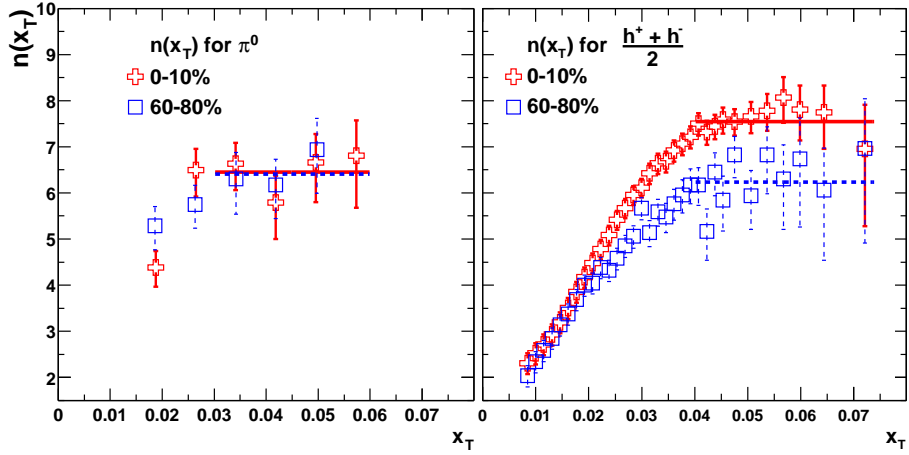}
\end{center}
	\vspace*{-0.24in}
\caption[]{Power-law exponent $n_{\rm eff}(x_T)$ for $\pi^0$ and $h$ spectra in central
and peripheral Au+Au collisions at $\sqrt{s_{NN}} = 130$ and 200 GeV \cite{Adler:2003au}. } 
\label{fig:nxTAA}
\end{figure}

If the production of high-$p_T$ particles in Au+Au collisions is the
result of hard scattering according to pQCD, then $x_T$ scaling should
work just as well in Au+Au collisions as in p-p collisions and should
yield the same value of the exponent $n_{\rm eff}(x_T,\sqrt{s})$.  The only assumption
required is that the structure and fragmentation functions in Au+Au
collisions should scale, in which case Eq. \ref{eq:bbg} still
applies, albeit with a $G(x_T)$ appropriate for Au+Au. In
Fig.~\ref{fig:nxTAA}, $n_{\rm eff}(x_T,\sqrt{s_{NN}})$ in Au+Au is shown for $\pi^0$ and $h^{\pm}$ in peripheral and central collisions, derived by
taking the ratio of $E d^3\sigma/dp^3$ at a given $x_T$ for
$\sqrt{s_{NN}} = 130$ and 200 GeV, in each case. 

The $\pi^0$'s exhibit
$x_T$ scaling, with the same value of $n_{\rm eff} = 6.3$ as in p-p collisions,
for both Au+Au peripheral and central collisions. The $x_T$ scaling establishes
that high-$p_T$ $\pi^0$ production in peripheral and central Au+Au
collisions follows pQCD as in p-p collisions, with parton distributions and 
fragmentation functions that scale with $x_T$, at least within
the experimental sensitivity of the data. The fact that the fragmentation functions scale for $\pi^0$ in Au+Au central collisions indicates that the effective energy loss must scale, i.e. $\Delta E(p_T)/p_T=$ is constant, which is  consistent with the parallel spectra on Fig.~\ref{fig:RAA}-(right) and the constant value of $R_{AA}$ as noted above. 
\subsection{The baryon anomaly}
\begin{figure}[!thb]
\begin{center}
\begin{tabular}{cc}
\includegraphics[width=0.52\linewidth]{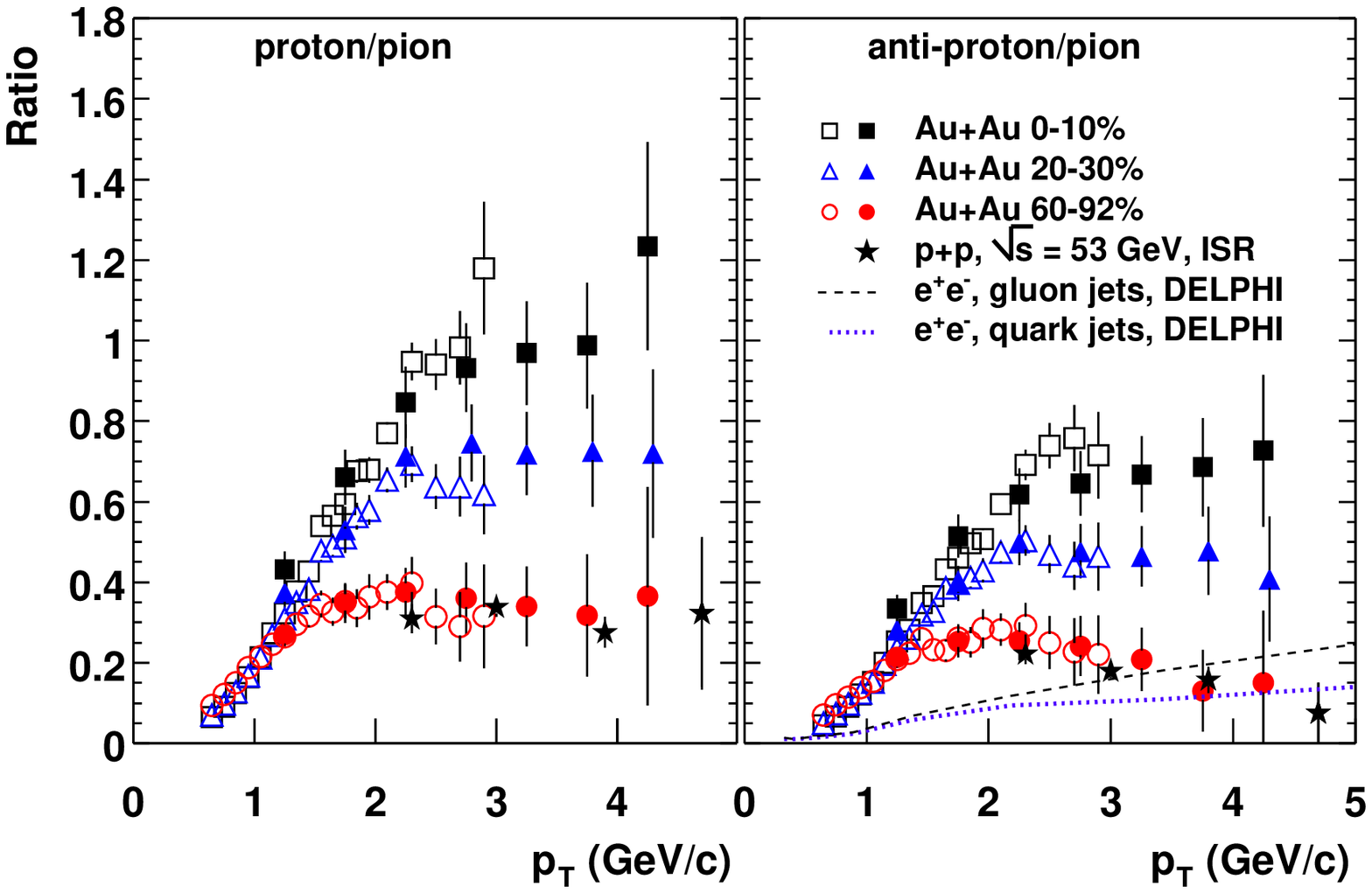}&
\includegraphics[width=0.40\linewidth]{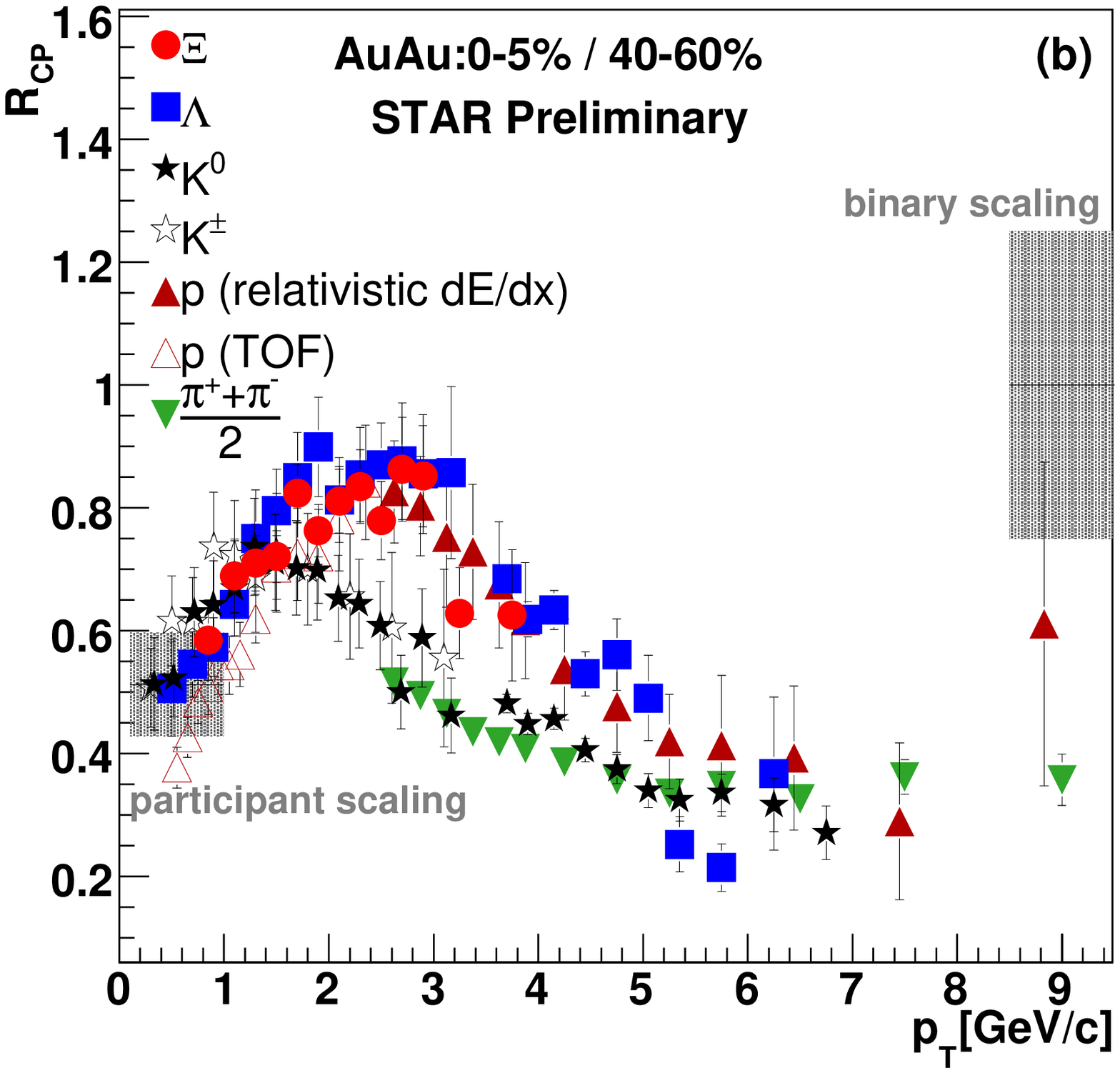}
\end{tabular}
\end{center}\vspace*{-0.25in}
\caption[]{(left) $p/\pi^+$ and $\bar{p}/\pi^-$ as a function of $p_T$ and centrality from Au+Au collisions at $\sqrt{s_{NN}}=200$ GeV~\cite{PXscalingPRL91}; (right) $R_{CP}$ for identified baryons and mesons~\cite{jana}.  }
\label{fig:banomaly}
\end{figure}

    The deviation of $h^{\pm}$ in Fig.~\ref{fig:nxTAA} from $x_T$ scaling in central Au+Au collisions is indicative of and consistent with the strong non-scaling modification of particle composition of identified hadrons observed in Au+Au collisions compared to that of p-p collisions in the range $2.0\leq p_T\leq 4.5$ GeV/c, where particle production is the result of jet-fragmentation. As shown in Fig.~\ref{fig:banomaly}-(left) the $p/\pi^{+}$ and $\bar{p}/\pi^{-}$ ratios as a function of $p_T$ increase dramatically to values $\sim$1 as a function of centrality in Au+Au collisions at RHIC~\cite{PXscalingPRL91} and $R_{CP}$ for identified $\pi^{\pm}$ mesons is much less than that for baryons (Fig.~\ref{fig:banomaly}-(right))~\cite{jana}. This is called the baryon anomaly, which is still not understood.  
Elegant models of coalescence of constituent quarks in a QGP~\cite{FMN,egMJTPoS06} have been proposed to explain the baryon anomaly at RHIC which predict a much larger effect at LHC. 
One such model~\cite{HwaYang} predicts that $p/\pi^+ \sim 10$ out to $p_T=20$ GeV/c at LHC from the recombination of partons from the many jets produced. Such a result, if observed at LHC, would be a major discovery. This does not involve a measurement of jets but requires single particle identification out to 20 GeV/c!
\subsection{Status of $R_{AA}$ at Quark Matter 2005} 
\begin{figure}[ht]
\begin{center}
\begin{tabular}{cc}
 \includegraphics[width=0.49\linewidth]{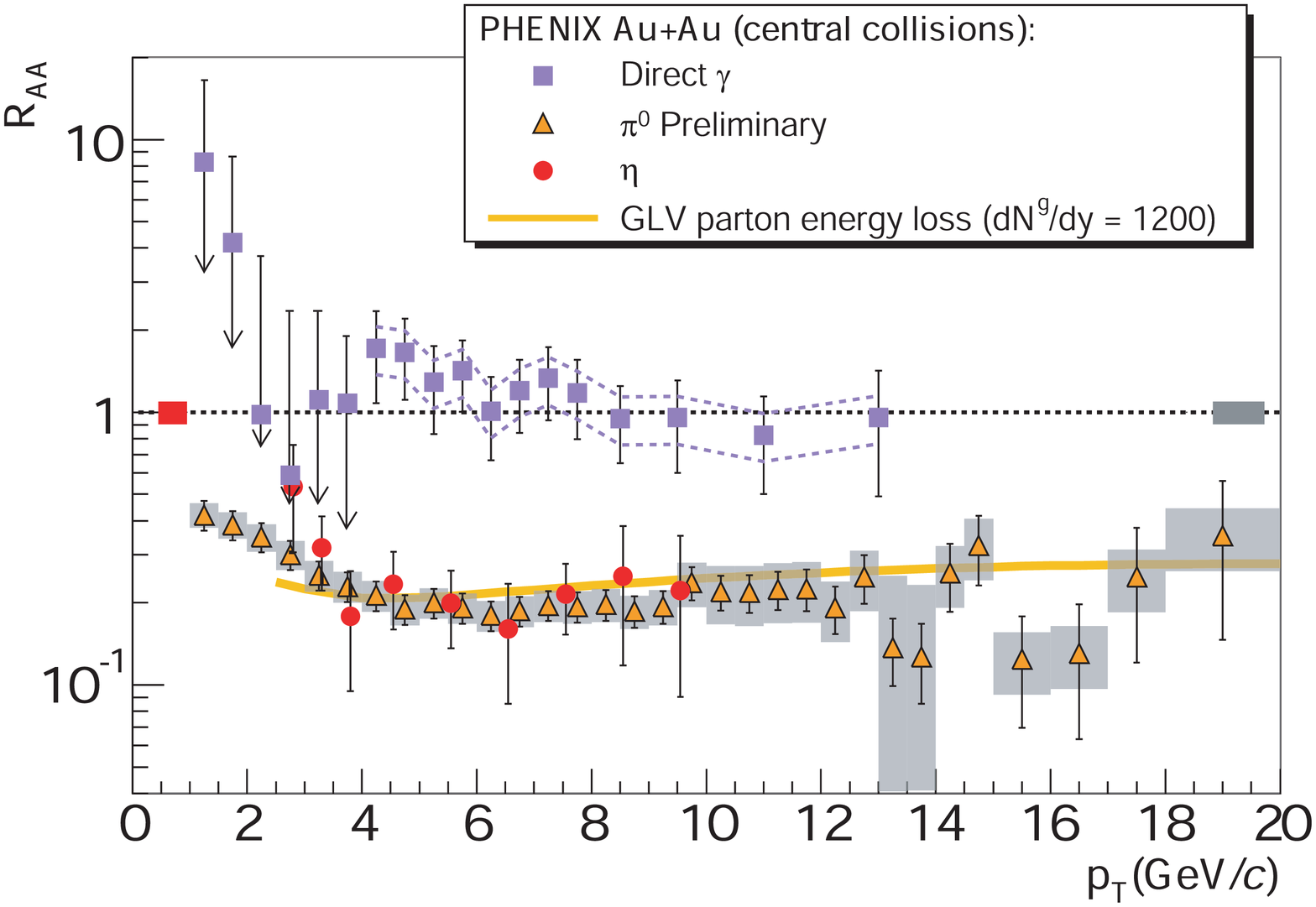}&
\hspace*{-0.02\linewidth}\includegraphics[width=0.49\linewidth]{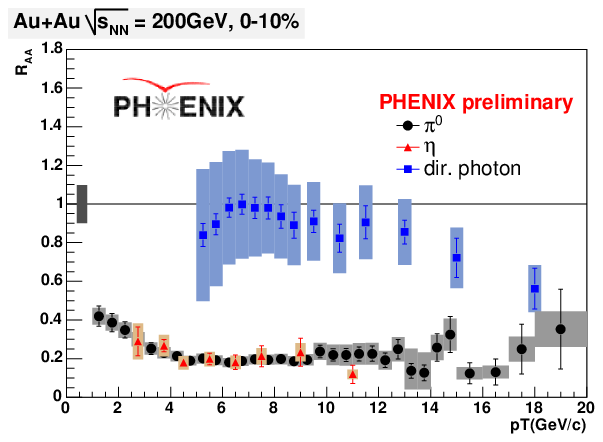}
\end{tabular}
\end{center}
\caption[]{(left) $R_{AA}(p_T)$ for direct-$\gamma$, $\pi^0$, $\eta$ vs. $p_T$ for Au+Au central (0-10\%) collisions at $\sqrt{s_{NN}}=200$ GeV~\cite{AkibaQM05}. (right) data from left with improved $R_{AA}(p_T)$ for direct-$\gamma$~\cite{BaldoQM06}. }
\label{fig:Tshirts} 
\end{figure}

  Single particle measurements painted a very clear picture of jet suppression at Quark Matter 2005~\cite{AkibaQM05} (Fig.~\ref{fig:Tshirts}-(left)). 
Direct $\gamma$ from the reaction $g+q\rightarrow \gamma+ q$ are not suppressed, while $\pi^0$ and $\eta$---which are both fragments of jets from final state partons---are suppressed by the same amount. This proves that the suppression is due to partonic interactions in the medium formed in central Au+Au collisions, since the direct-$\gamma$'s escape from the medium without interacting while the initial partonic composition is not much different for direct-$\gamma$ and $\pi^0$ production. It was also comforting to some to see a theoretical curve~\cite{GLV} from the GLV model of partonic energy loss in the medium lie nearly on the data. However, the GLV model~\cite{GLV} appears to show an increase of $R_{AA}$ for $\pi^0$ with increasing $p_T$  while the data appear flat to within the errors which clearly could still be improved in the range $12\leq p_T\leq 20$ GeV/c. 

   If we experimenters had stopped here, many people would have been happy and claimed that RHIC physics was understood~\cite{butcharm}, it was time to move to LHC: i.e. skim the physics from $\sqrt{s_{NN}}\sim 5$ to 5500 GeV, a factor of $> 1000$ in $\sim 20$ years (!), without really understanding it in detail.  They were wrong in 2005 and are certainly just as wrong now. 
   
\begin{figure}[ht]
\begin{center}
\begin{tabular}{cc}
\includegraphics[width=0.46\linewidth]{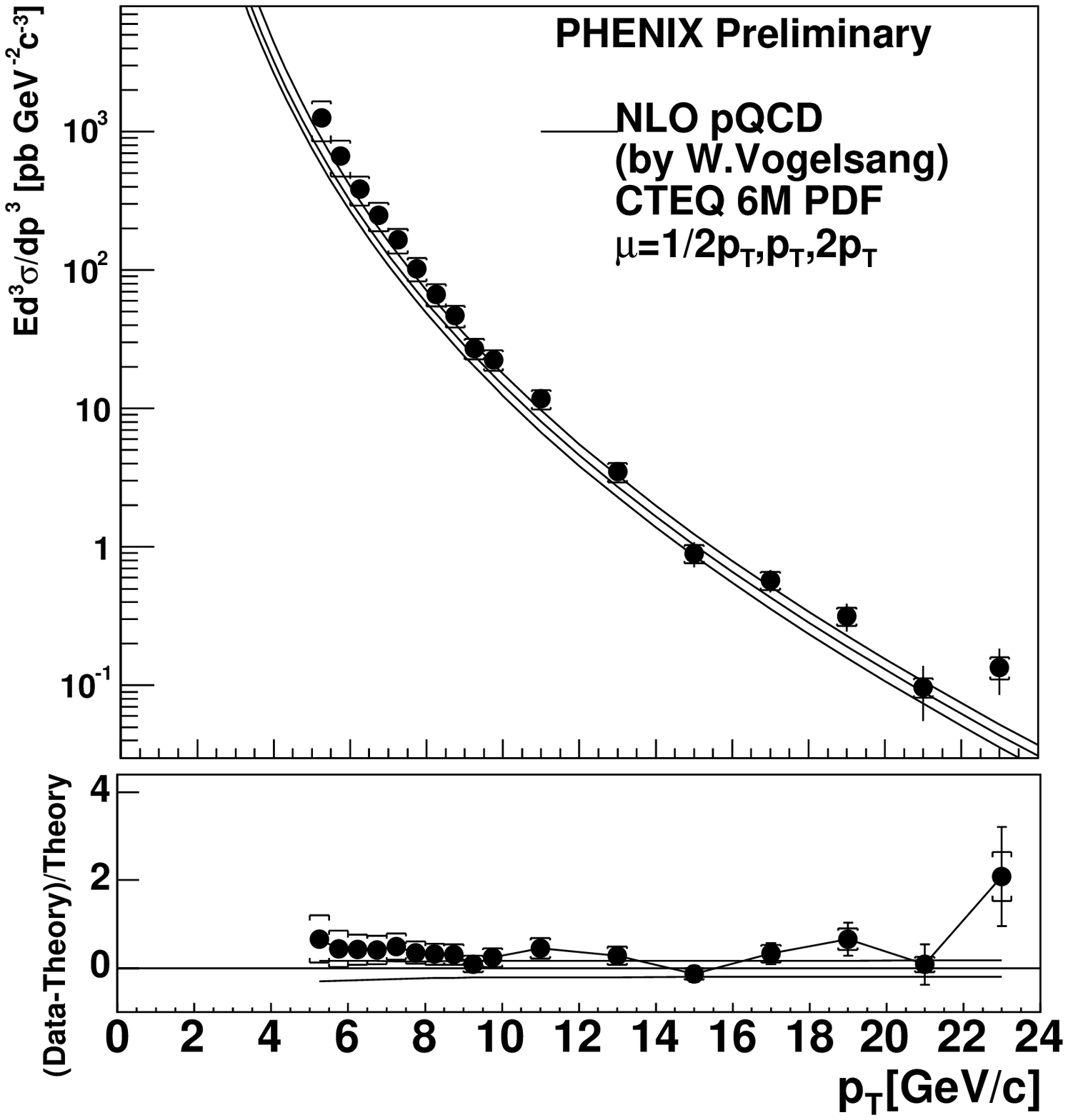}&
\includegraphics[width=0.47\linewidth]{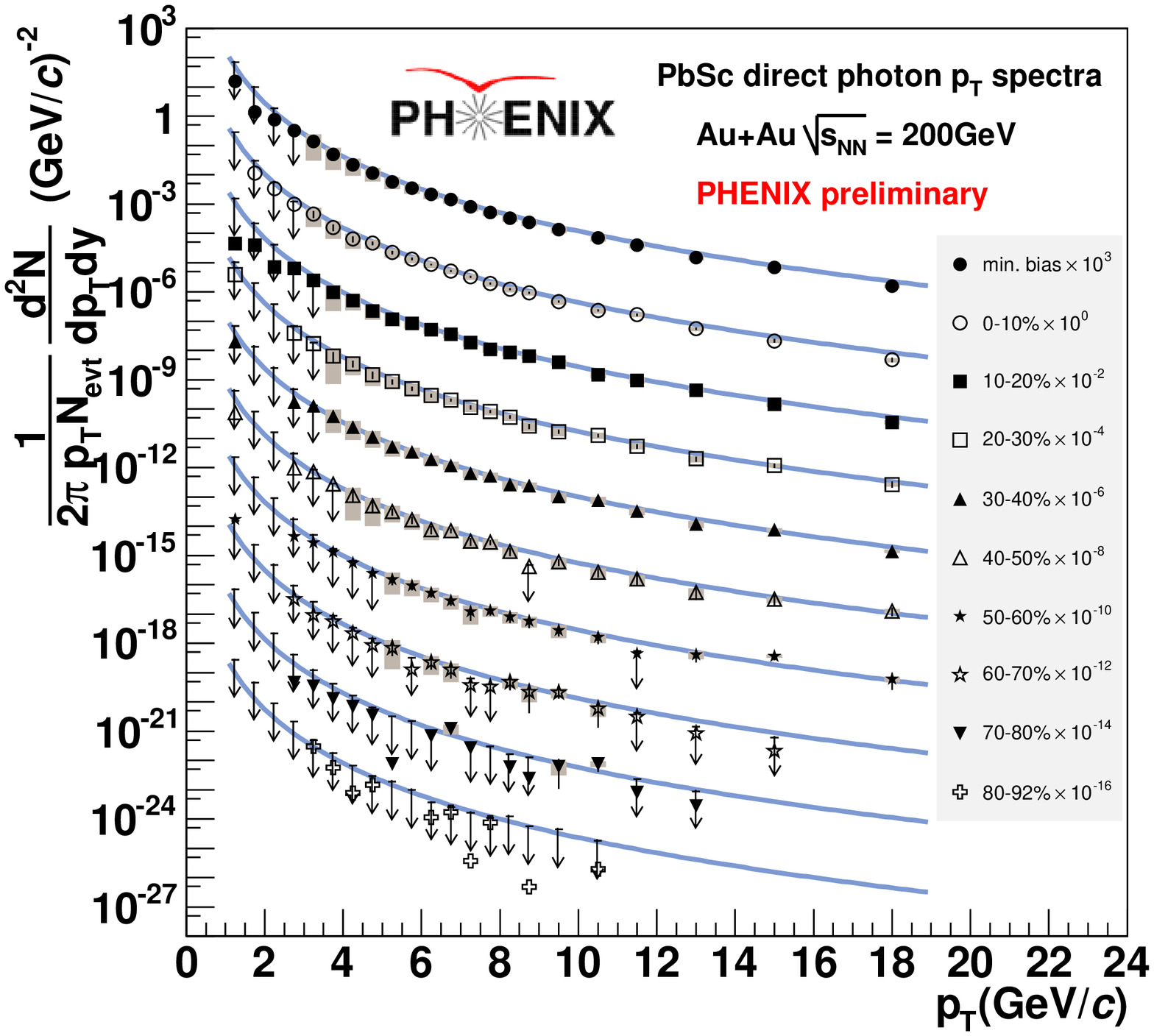}
\end{tabular}
\end{center}
\caption[]{PHENIX direct-$\gamma$ measurements~\cite{IsobeQM06} at $\sqrt{s_{NN}}=200$ GeV. (left) p-p collisons, (right) Au+Au collisions vs centrality. Solid lines are pQCD (scaled by $\mean{T_{AA}}$).}
\label{fig:QM06photon} 
\end{figure}

   The improvement of the direct-$\gamma$ measurement in both p-p and Au+Au collisions by Quark Matter 2006~\cite{IsobeQM06} (Fig.~\ref{fig:QM06photon}) led to an extension of $R_{AA}$ for direct-$\gamma$ to $p_T=20$ GeV/c with astounding results. 
   It appeared from the preliminary data (Fig.~\ref{fig:Tshirts}-(right)) that the direct-$\gamma$ were becoming suppressed as much as the $\pi^0$ (to within the systematic errors) as $p_T\rightarrow 20$ GeV/c. If this were true, i.e. $R^{\pi^0}_{AA}=R^{\gamma}_{AA}$ for $p_T > 20$ GeV/c, it would indicate that the suppression in this higher $p_T$ range is not a final state effect (the energy loss has become negligible compared to $p_T\sim 20$ GeV/c) and therefore must be an initial state effect due to the structure functions. This would have a dramatic effect on LHC physics since it would imply that $\sim 60$ GeV jets at LHC likely would be beyond the range where medium effects would be apparent, hence pointless to study for RHI physics.

\section{Direct photon production and structure functions}
   Direct photon production is one of the best reactions to study QCD in hadron collisions, since there is direct and unbiased access to one of the interacting constituents, the photon, which can be measured to high precision, and production is predominantly via a single subprocess  
\begin{equation}
g + q\rightarrow \gamma + q \;\;\;\;\; ,
\label{eq:QCDcompton}
\end{equation}
with $q + \bar q\rightarrow \gamma + g$ contributing on the order of 10\%. 
   However, the measurement is relatively difficult experimentally due to the huge  background of photons from $\pi^0\rightarrow \gamma+\gamma$ and $\eta\rightarrow \gamma+\gamma$ decays. Clearly the beautiful measurements of direct-$\gamma$ at RHIC in both p-p and Au+Au collisions show that this problem can be overcome. Direct-$\gamma$ production is a crucial measurement to be made at LHC, likely one of the most important measurements. 
   
   In light of the fact that the reaction is dominated by one subprocess (Eq.~\ref{eq:QCDcompton}), the minimum bias cross section for direct-$\gamma$ production at mid-rapidity, apart from kinematic factors which cancel in the ratios p+A/p-p or A+A/p-p, depends only on the initial state structure functions. 
   For p+A collisions, $R^{\gamma}_{pA}$ measures the average of the quark and gluon structure function ratios:    
   \begin{equation}
   R^{\gamma}_{pA}(p_T)={{d^2\sigma^{\gamma}_{pA}(p_T)/dp_T dy} \over {A\times d^2\sigma^{\gamma}_{pp}(p_T)}/dp_T dy}\approx 
   {1\over 2}\left({ {F_{2A}(x_T)} \over {A F_{2p}(x_T)}}+{ {g_{A}(x_T)} \over {A g_{p}(x_T)}}\right)={1\over 2}\left[R^A_{F_2}(x_T,Q^2)+R^A_g(x_T,Q^2)\right] 
   \label{eq:RgammapA}
   \end{equation}
   where ${F_{2A}(x)/AF_{2p}(x)}\equiv R^A_{F_2}(x,Q^2)$ is the ratio of the $F_2$ structure functions from $e(\mu)-A$ and $e(\mu)-p$ Deeply Inelastic Scattering (DIS), and ${g_{A}(x)/Ag_p(x)}\equiv R^A_g(x,Q^2)$ is the ratio of the gluon structure function in a nucleus $A$ to that in a proton. For A+A collisions, $R^{\gamma}_{AA}$ measures the product of the quark and gluon structure function ratios: 
     \begin{equation}
   R^{\gamma}_{AA}(p_T)={{d^2\sigma^{\gamma}_{AA}(p_T)/dp_T dy} \over {A^2\times d^2\sigma^{\gamma}_{pp}(p_T)}/dp_T dy}\approx 
   { {F_{2A}(x_T)} \over {A F_{2p}(x_T)}}\times{ {g_{A}(x_T)} \over {A g_{p}(x_T)}}=R^A_{F_2}(x_T,Q^2)\times R^A_g(x_T,Q^2)\;. 
   \label{eq:RgammaAA}
   \end{equation} 
   
\begin{figure}[!thb]
\begin{center}
\begin{tabular}{cc}

\includegraphics[width=0.48\linewidth]{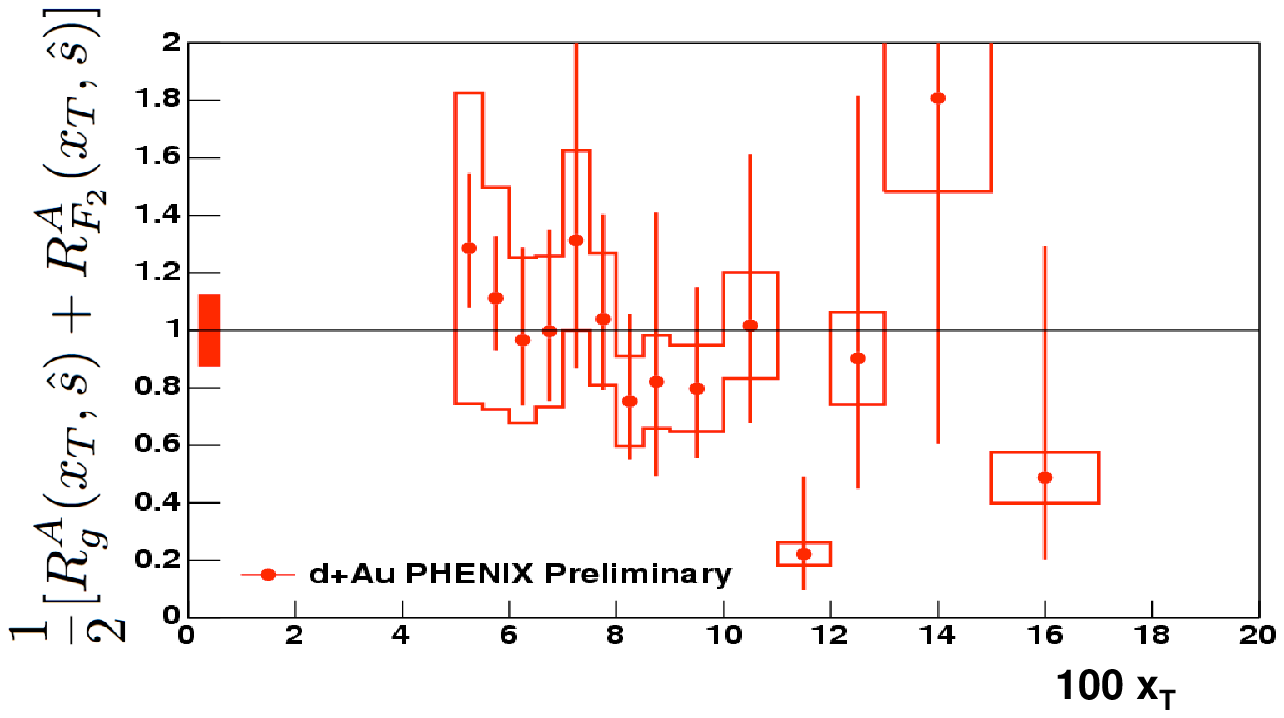}&
\includegraphics[width=0.48\linewidth]{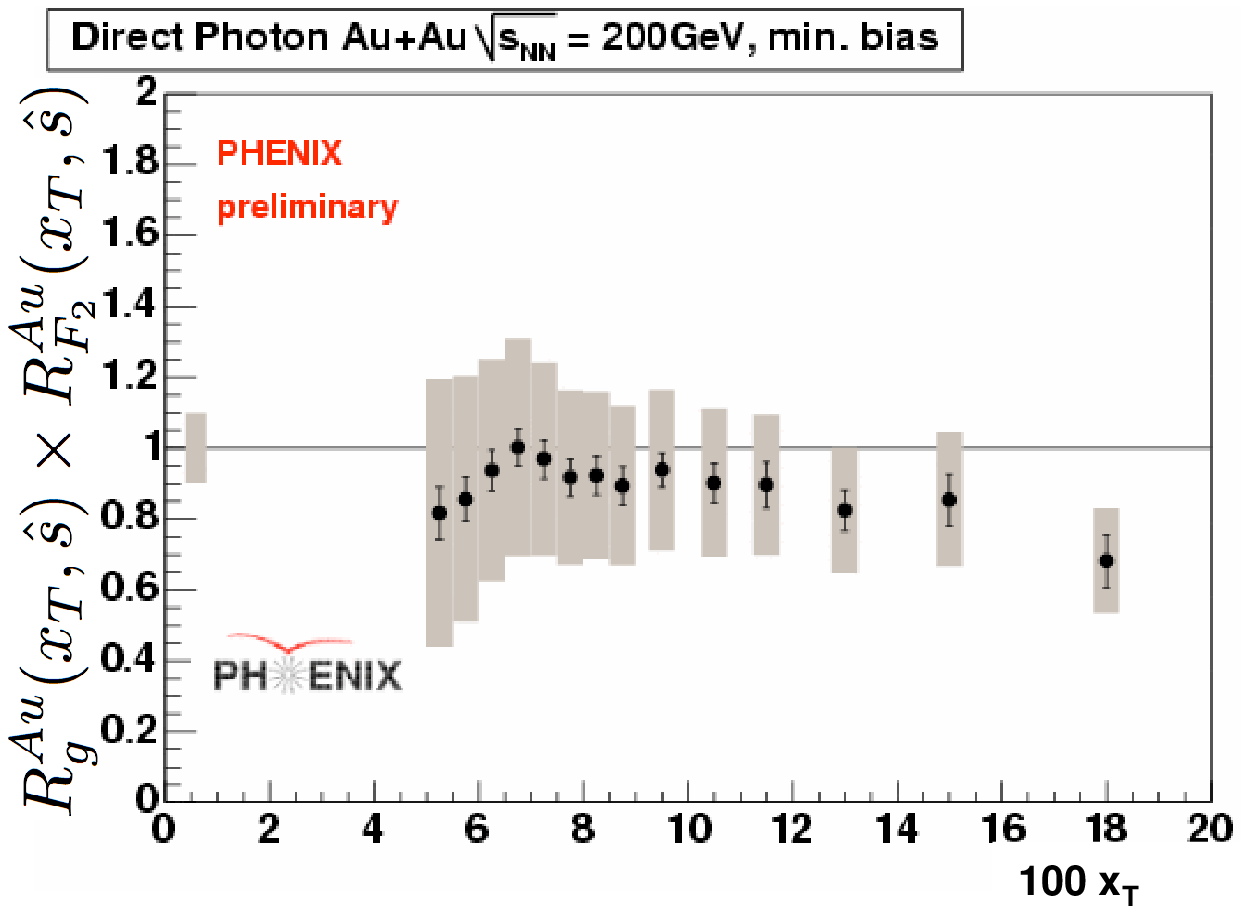}

\end{tabular}
\end{center}\vspace*{-0.25in}
\caption[]{(left) $R^{\gamma}_{dAu}(p_T)$~\cite{PeressHP06}, (right) $R^{\gamma}_{AuAu}(p_T)$~\cite{IsobeQM06} in minimum-bias collisions at $\sqrt{s_{NN}}=200$  GeV.  }
\label{fig:PXdAAA}

\end{figure}

	The PHENIX data for $R_{dA}$ and $R_{AA}$ minimum bias measurements labelled according to Eqs.~\ref{eq:RgammapA}, \ref{eq:RgammaAA} are shown in Fig.~\ref{fig:PXdAAA} 
and may be directly compared to measurements or predictions of the structure function ratios in the range $0.05\leq x\leq 0.20$~\cite{ispin}. For central Au+Au collisions, $R^{\gamma}_{AA}$ (Fig.~\ref{fig:Tshirts}-(right)) measures the structure function ratios for central collisions. However, one can not check these by comparing to experimental data because in the 40 years since DIS was discovered, no experiment has ever measured a structure function as a function of centrality~\cite{noteE665}. Measurements of $R^A_g$ as a function of centrality can be made in p+A collisions at RHIC; while $R^A_{F_2}$ could be measured at eRHIC, with the challenge being to invent a method to measure centrality in $e-A$ collisions. Note that since $R_{AA}\propto AA/pp=(A/p)^2$,  isotopic spin issues with d+Au data can be avoided by using p+A comparison data in the future. 
\subsection{Direct photons at LHC}
\begin{figure}[!b]
\begin{center}
\begin{tabular}{cc}
\includegraphics[width=0.48\linewidth]{figs/YellowgluonEMC.epsf}&
\includegraphics[width=0.45\linewidth]{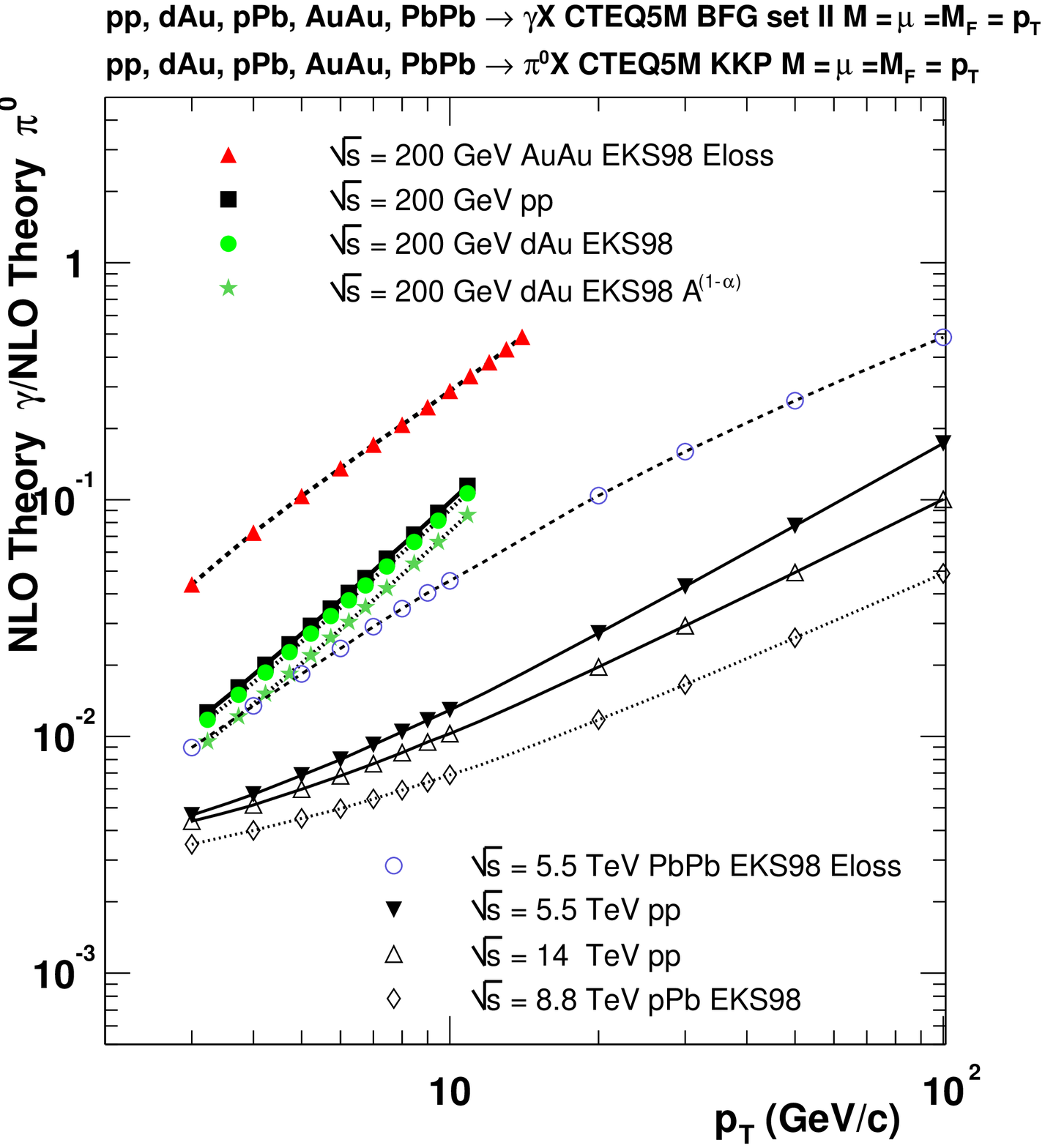}
\end{tabular} 
\end{center}
\caption[]{(left) Predictions for $R_g^{Pb}(x)$~\cite{CY2004-009-1}. (right) Predictions for direct-$\gamma/\pi^0$~\cite{CY2004-009-4}. }
\label{fig:dirgLHC} 
\end{figure}

  Single particle inclusive measurements in Pb+Pb collisions at LHC in the range $2\leq p_T\leq 20$ GeV/c ($0.0007\leq x_T\leq 0.007$) are of prime importance  but they may be difficult to understand without comparison data in p-p and p+Pb collisions because the structure functions in nuclei are unknown in this $x$ range. The theoretical estimates vary by large factors (see Fig.~\ref{fig:dirgLHC}-(left)).  
This is where a direct-$\gamma$ measurement can be most informative. However, direct-$\gamma$ measurements are much more difficult in the region of interest at the LHC than at RHIC because the $\gamma/\pi^0$ at a given value of $p_T$ is much smaller (Fig.~\ref{fig:dirgLHC}-(right)), so that the background is larger. Also, since hard-scattering has a strong dependence on $\sqrt{s_{NN}}$ (Eq.~\ref{eq:bbg}), it is imperative that p-p comparison data be taken at $\sqrt{s}=\sqrt{s_{NN}}=5500$ GeV for the required precision in $R_{AA}$. 

\begin{figure}[ht]
\begin{center}
\begin{tabular}{cc}
\includegraphics[width=0.60\linewidth]{figs/v2gamma.epsf}&\hspace*{-0.04\linewidth}
\includegraphics[width=0.38\linewidth,height=0.40\linewidth]{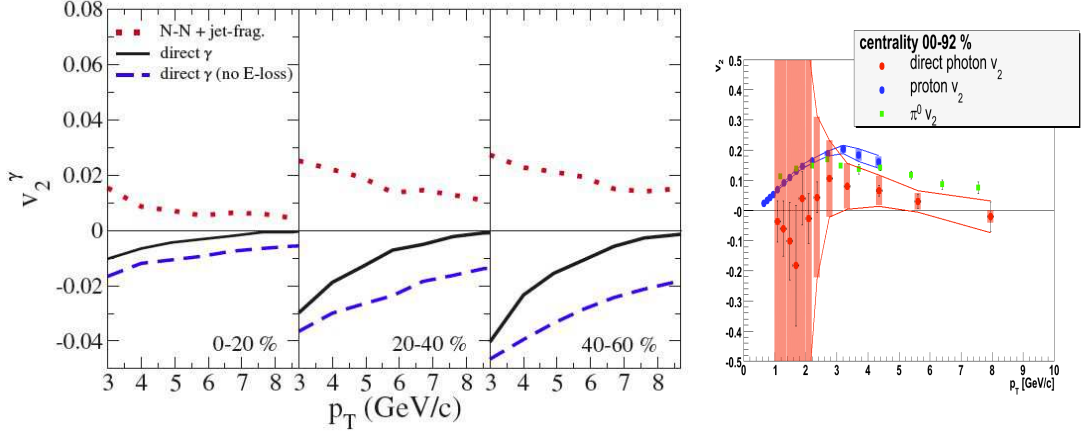}
\end{tabular} 
\end{center}
\caption[]{(left) Predictions~\cite{TGF} of $v_2(p_T)$ of photons from various sources as labeled for 3 values of centrality for Au+Au collisions at RHIC. (right) PHENIX preliminary~\cite{MikiQM06} values of $v_2(p_T)$ for direct-$\gamma$, $p$ and $\pi^0$ in Au+Au minimum bias collisions at $\sqrt{s_{NN}}=200$ GeV.}
\label{fig:v2gamma} 
\end{figure}

	Should the LHC physicists be so lucky that direct-$\gamma$ are not suppressed in Pb+Pb collisions in the region of interest (i.e. all the theoretical estimates in Fig.~\ref{fig:dirgLHC}-(left) are wrong), the next thing to check would be the azimuthal anisotropy of direct-$\gamma$ production with respect to the reaction plane (represented by the second fourier component $v_2$) because there is a prediction~\cite{TGF} that if jet suppression is due to the partonic interaction $q+g\rightarrow q+g +g$ in the medium, then the process $q+g\rightarrow \gamma+q$ will also exist which will generate a source of direct-$\gamma$ from the medium. The generated $\gamma$ will have the opposite sign of $v_2$ compared to fragments of partons, since more partons will be absorbed and more photons created along the long axis of the almond-shaped region of nuclear overlap, with less in each case for the short axis (see Fig.~\ref{fig:v2gamma}-(left)). So far at RHIC there is no evidence for such an effect (Fig.~\ref{fig:v2gamma}-(right)), as the $v_2$ of direct-$\gamma$ is consistent with $v^{\gamma}_2(p_T)=0$ within the systematic error, and more likely to be the same sign rather than the opposite sign of $v^{\pi^0}_2$.    

	At LHC there is the interesting possibility of making this measurement with low mass $e^+ e^-$ and $\mu^+ \mu^-$ pairs from internal and external conversions of photons with $p_T\lsim 5$ GeV/c. The $e^+ e^-$ pairs are predominantly from conversions of photons from $\pi^0\rightarrow \gamma +\gamma$ so should represent the hadronic $v_2$, while the $\mu^+ \mu^-$ pairs have rigorously zero $\pi^0$-Dalitz contribution and minimal $\eta$-Dalitz contribution so should represent the direct-$\gamma$ $v_2$ since the external conversions are down by a factor of $m_{\mu}^2/m_{e}^2\approx 40,000$. Opposite $v_2$ for $e^+ e^-$ pairs and $\mu^+ \mu^-$ pairs would indicate dramatic new physics without need for comparison runs or detailed knowledge of the background.
 
    Should $R^{\gamma}_{AA}=1$ and $v^{\gamma}_2$=0 at LHC, one could consider skipping the p+Pb (or d+Pb) comparison runs at $\sqrt{s_{NN}}=5500$ GeV, which 
will be more difficult to interpret than at RHIC due to the moving c.m. system (or the totally unknown, possibly unknowable, isotopic spin corrections). 
     
\section{Jets via 2-particle correlations.}
The huge multiplicity in central A+A collisions makes jet reconstruction difficult if not impossible at RHIC. However, two-particle correlations work well, with the major complication being the event-anisotropy ($v_2$) with respect to the reaction plane defined by the impact parameter vector in Au+Au collisions.  
  \begin{figure}[!ht]
\begin{center}
\begin{tabular}{cc}
\begin{tabular}[b]{c}
\hspace*{-0.03\linewidth}\includegraphics[width=0.451\linewidth]{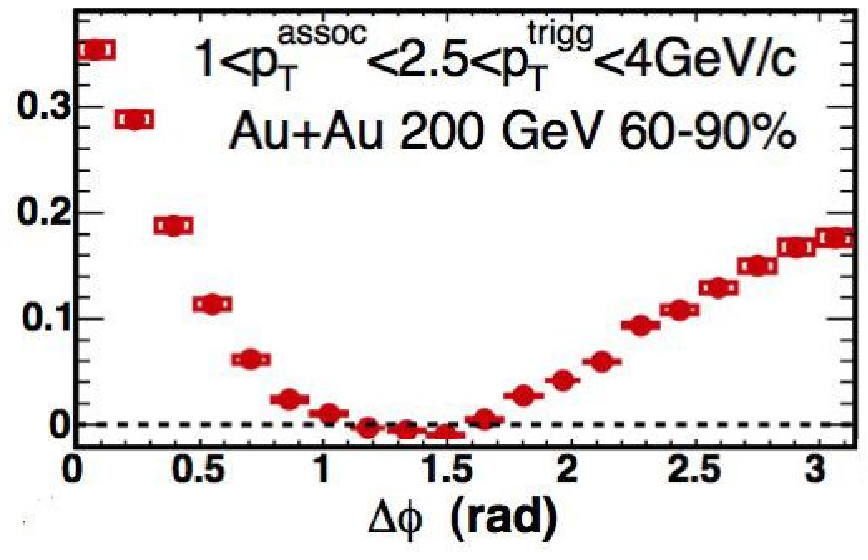}\hspace*{0.03\linewidth}\cr
\hspace*{-0.03\linewidth}\includegraphics[width=0.528\linewidth]{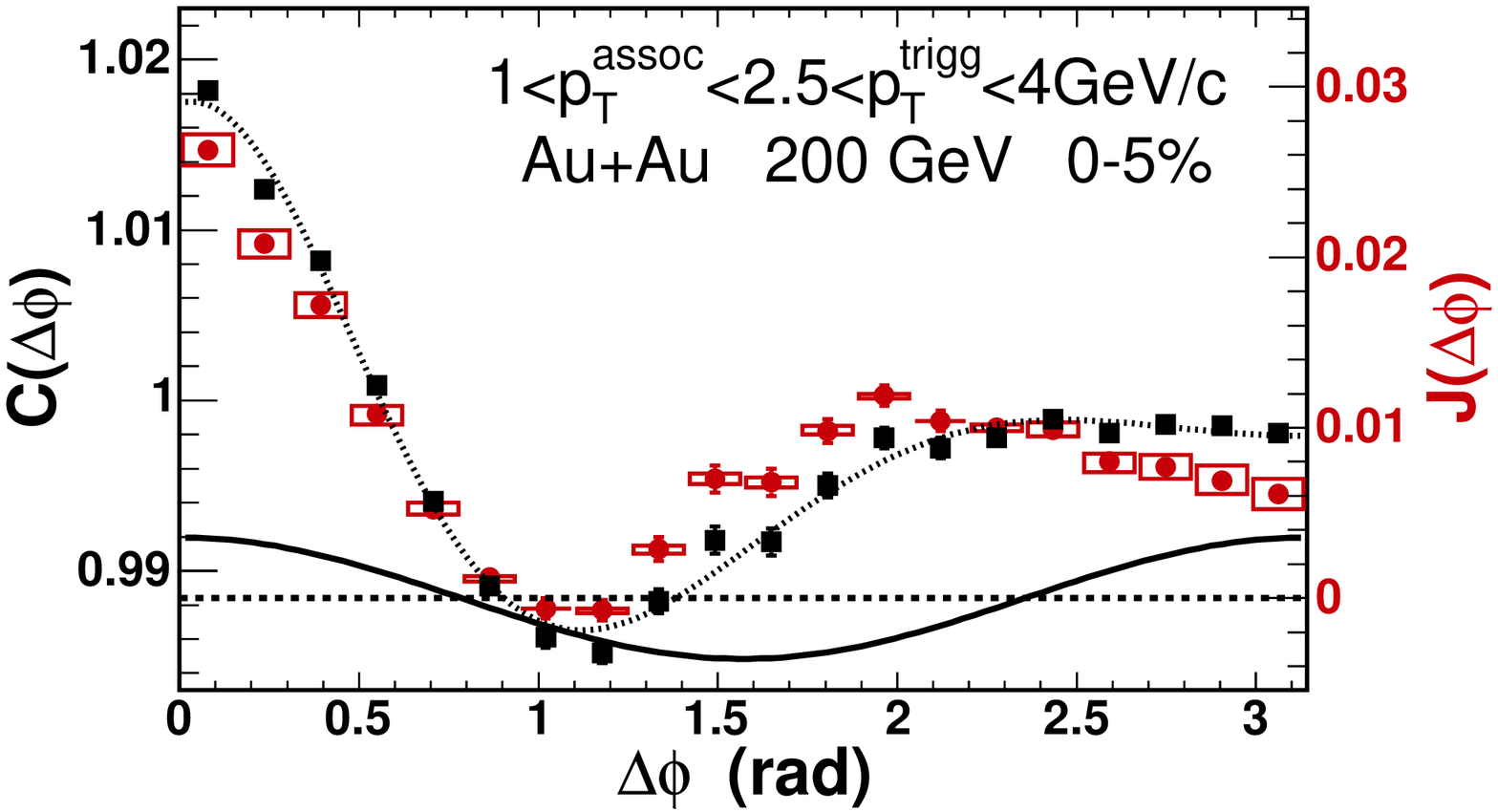}
\end{tabular} 
\hspace*{-0.001\linewidth}\includegraphics[width=0.48\linewidth]{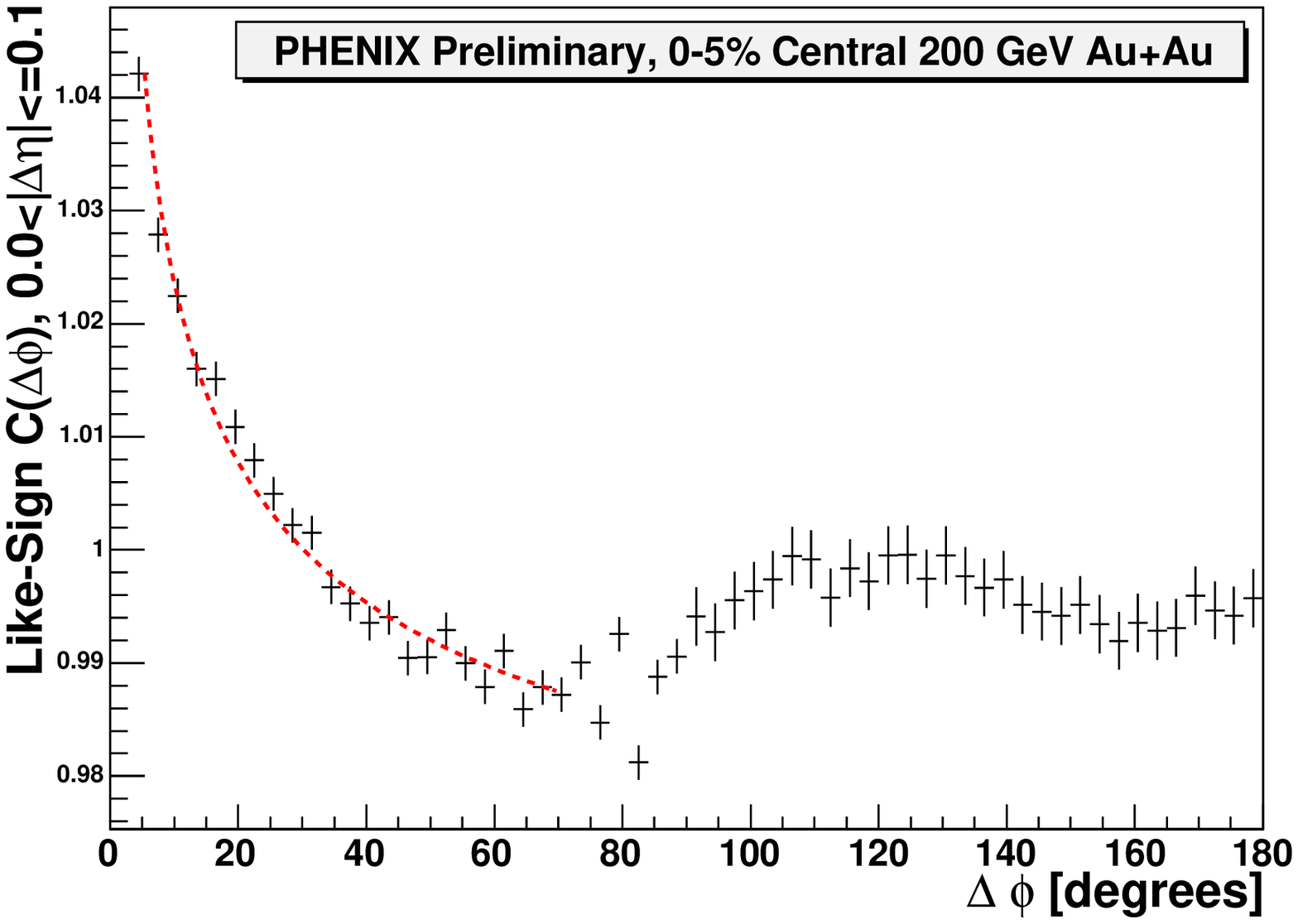} 
\end{tabular}
\end{center}
\caption[]{(left) Azimuthal correlation $C(\Delta\phi)$ of $h^{\pm}$ with $1\leq p_{T_a}\leq 2.5$ GeV/c with respect to a trigger $h^{\pm}$ with $2.5\leq p_{T_t} \leq 4$ GeV/c in Au+Au~\cite{ppg067}: (top) central collisions, where the line with data points indicates $C(\Delta\phi)$ before correction for the azimuthally modulated ($v_2$) background, and the other line is the $v_2$ correction which is subtracted to give the jet correlation function $J(\Delta\phi)$ (data points); (bottom)-same for peripheral collisions. (right) Auto-correlation of like-sign pairs of $0.2 < p_T <0.4$ GeV/c particles in central Au+Au collisions~\cite{JTMQM06}.}
\label{fig:f8}
\end{figure}
In Fig.\ref{fig:f8}-(left)~\cite{ppg067}, even before the $v_2$ subtraction, the away-side distribution in central collisions is much wider than in peripheral collisions (which is similar to p-p). After the $v_2$ correction, a dip develops at 180$^\circ$ leading to a double peak structure $\sim \pm 1$ radian from 180$^\circ$. The double peak persists to low values of $p_{T_t}$, $p_{T_a}~\sim 0.3$ GeV/c which may indicate that it is a reaction of the medium to the passage of hard-scattered partons~\cite{JTMQM06}.  
STAR has also reported the widening of jet-correlations in the ``intermediate'' $p_T$ range ($2\leq p_{T_t}\leq 4.5$ GeV/c) as well as other new phenomena such as the ``ridge'', and punch-through of seemingly unscathed jets (apparently, with the same width as in p-p collisions) for away-side tracks with $p_{T_a}\geq 3$ GeV/c from charged hadron triggers with $8< p_{T_t} < 15$ GeV/c~\cite{egMJTPoS06}. A debate `rages' between STAR and PHENIX about whether the wide jets are evidence of large deflection of jets in the medium, or whether the double peak is evidence of a conical reaction to the medium such as a ``Mach Cone''~\cite{Mach}. Studies of these effects as a function of the reaction plane are just beginning. 
\section{Conclusions}
  Frankly, until all the new, exciting and unexplained results at RHIC for $0.1 < p_T < 20$ GeV/c are understood, I am not interested in $\sim 100$ GeV jets at LHC nor do I expect much influence of the medium to be observable at such large $p_T$. On the positive side, I have mentioned several possible identified single particle measurements at LHC which I believe are much more likely to reveal new physics. 
  
\begin{figure}[ht]
\begin{center}
\begin{tabular}{cc}
\includegraphics[width=0.40\linewidth]{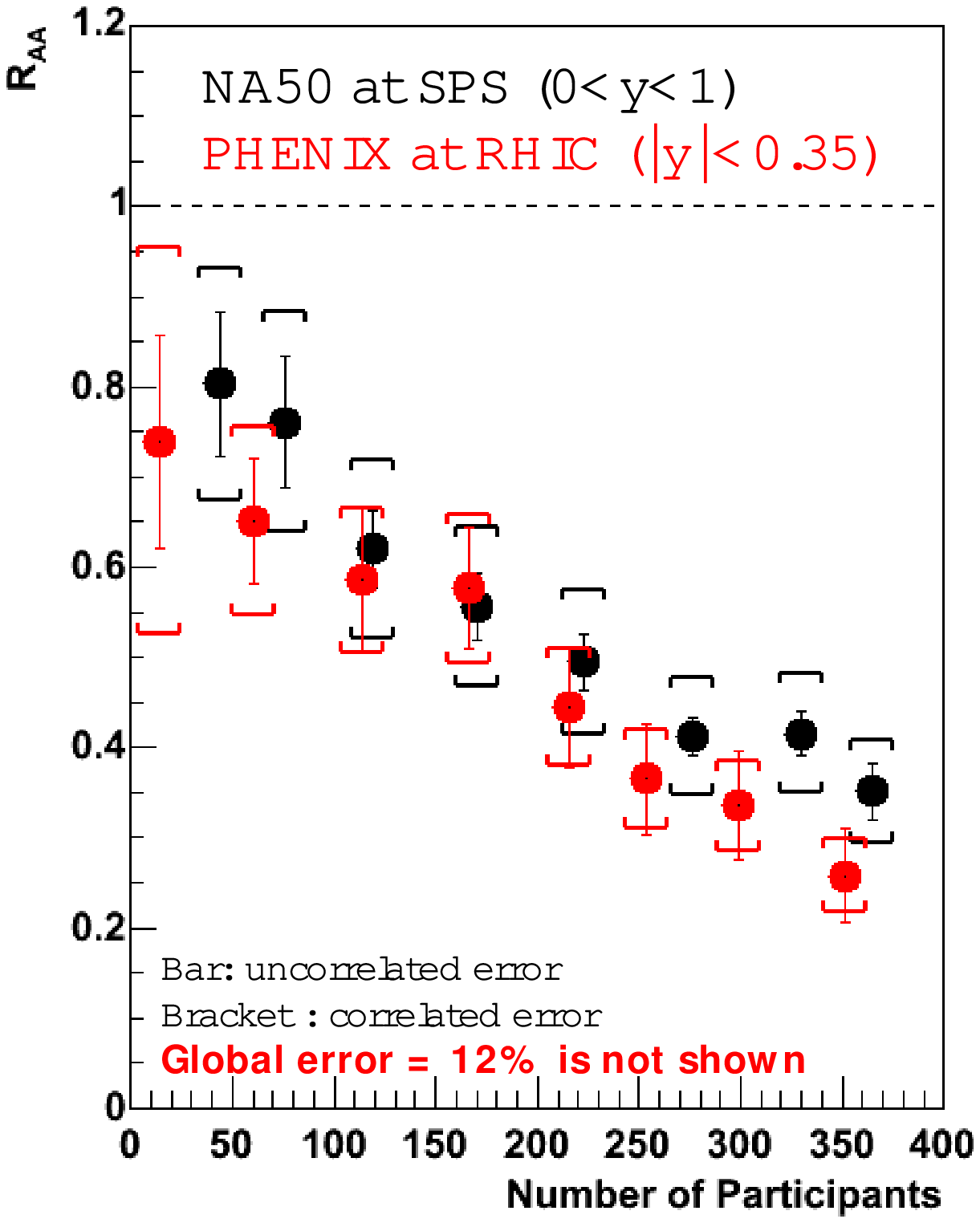}&
\includegraphics[width=0.48\linewidth]{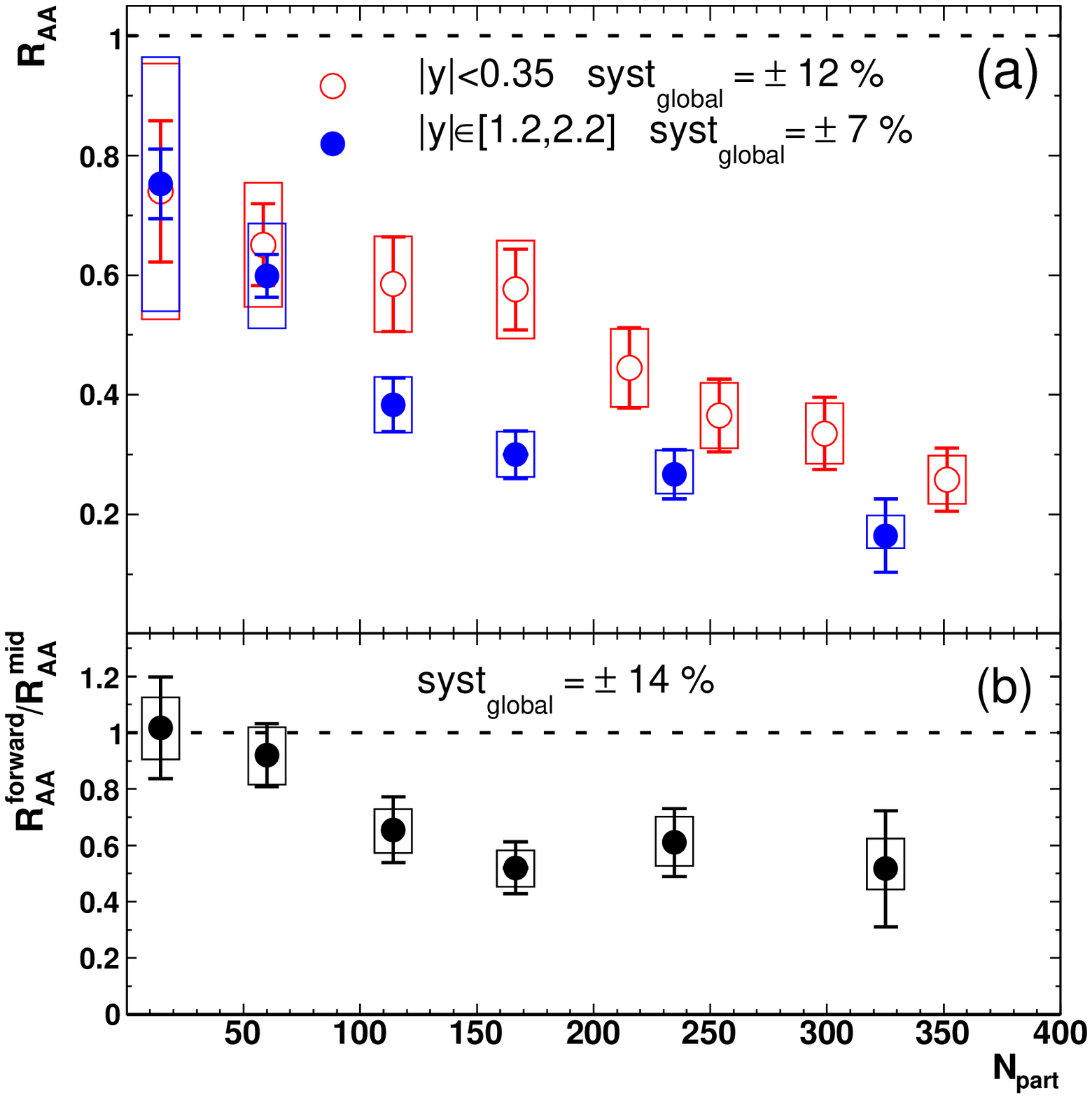} 
\end{tabular}
\end{center}
\caption[]{(left) $R_{AA}$ for $J/\Psi$ at mid-rapidity as a function of centrality~\cite{GunjiQM06}: PHENIX (Au+Au, $\sqrt{s_{NN}}=200$ GeV); NA50 (Pb+Pb, $\sqrt{s_{NN}}=17.2$ GeV). (right) $R_{AA}$ for $J/\Psi$ as a function of centrality in two rapidity intervals for $\sqrt{s_{NN}}=200$ GeV Au+Au collisions~\cite{ppg068}.  }
\label{fig:JPsi} 
\end{figure}

  After 6 runs at RHIC, many discoveries have been made in Au+Au collisions but there is much that is still not known or understood:
  \begin{itemize}
\item Is the nuclear modification factor $R_{AA}$ for $\pi^0$ really constant at a factor of 5 suppression over the range $3< p_T< 20$ GeV/c, which would occur for a constant-fractional energy loss analogous to bremsstrahlung, or does the suppression tend to vanish at larger $p_T$? Is $dE/dx$ constant or a constant fraction or something else?
\item Does $R_{AA}$ for direct-$\gamma$ really approach that of $\pi^0$ at large $p_T \sim 20$ GeV/c as indicated by preliminary data? If true, this would argue that the suppression due to a medium effect vanishes at large $p_T > 20$ GeV/c and the remaining suppression is due to the structure functions. If this is confirmed, it would be very bad for the LHC.
\item The detailed mechanism of jet suppression due to interaction with the medium is not understood. It is not known whether partons lose energy continuously or discretely, whether they stop in the medium so that the only observed jet fragments are those emitted from the surface or whether partons merely lose energy exiting the medium such that those originating from the interior of the medium with initially higher $p_T$ are submerged (due to the steeply falling $p_T$ spectrum) under partons emitted at the surface which have not lost energy. In either case, there is a surface bias. 
\item If the jet punch-through is due to tangential jet-pairs, why does it depend so strongly on $p_{T_t}$?
\item It is not known whether a parton originating at the center of the medium can exit the medium without losing energy. 
\item It is not known where the energy from the absorbed jets or parton energy loss goes or how it is distributed. 
\item The reason why heavy quarks appear to lose the same energy as light quarks is not understood. 
\item The surface bias discussed above complicates the use of two-particle correlations of hard-scattered partons to probe the medium since detecting a particle from an away-side parton changes the surface bias of the trigger parton. This means that detection of both a trigger and away side particle is required in order to constrain the hard-scattering kinematics and the position of the origin of the hard-scattered parton-pair within the nuclear matter. Then, the main correlation information with relatively stable kinematics and origin is obtained by studying correlations with an additional 1 or two particles, i.e. a total of 3 or 4 particle correlations, which is much more complicated and requires much more data than the same studies in in p-p collisions. 
  \item The baryon anomaly, the increase of the $p^{\pm}/\pi^{\pm}$ ratio in the range $2<p_T <6$ GeV/c in Au+Au collisions from the value given by parton-fragmentation in this $p_T$ range in p-p collisions, is not understood. Elegant recombination models fail to explain the similar jet activity  correlated to the $p^{\pm}$ and $\pi^{\pm}$ triggers in this ``intermediate'' $p_T$ range. 
  \item The wide away-side non-identified hadron correlations for triggers in the intermediate range $2<p_{T_t} <6$ GeV/c in Au+Au collisions, with a possible dip at $180^\circ$ which causes apparent peaks displaced by  $\sim 60^\circ$, is not understood. It could represent a Mach cone due to the analogy of a sonic-boom of the parton passing through the medium faster than the speed of sound, or it could indicate jets with large deflections. The effect may be related to the baryon anomaly, which occurs in this $p_T$ range; or the peaks, which are seen also for much softer trigger particles, may not be a hard-scattering effect. 
\item The ridge is not understood. What causes it? What are its properties?  How does it depend on $p_{T_t}$, angle to reaction plane etc? Why isn't there an away-side ridge?
\item Finally, $J/\Psi$ suppression, which for more than 20 years has represented the gold-plated signature of deconfinement, is not understood (see Fig.~\ref{fig:JPsi}).  
\end{itemize}


\begin{thebibliography}{99}
\bibitem{ispin} $R_{dA}$ must be corrected for isotopic spin effects to obtain $R_{pA}$. This was not done in Fig.~\ref{fig:PXdAAA}.
\end{thebibliography}
\end{document}